\documentclass{article}
\usepackage[utf8]{inputenc}
\usepackage[T1]{fontenc}
\usepackage{lineno}
\usepackage{bm}
\usepackage{csquotes}
\usepackage{float}
\usepackage{longtable}
\usepackage{array}

\usepackage{PRIMEarxiv}

\usepackage{hyperref}       
\usepackage{url} 
\usepackage{booktabs} 
\usepackage{todonotes}
\usepackage[ruled,vlined]{algorithm2e}
\usepackage{adjustbox}
\usepackage{listings}%
\usepackage{algpseudocode}%
\usepackage{algorithmicx}%
\usepackage{graphicx}
\usepackage{subcaption}
\usepackage{multirow}
\usepackage{amsmath}
\usepackage{makecell}
\usepackage{graphics}
\usepackage{tcolorbox}
\usepackage{xcolor}
\usepackage{enumitem}

\definecolor{nb}{HTML}{006EB8}
\usepackage{hyperref}
\hypersetup{
  colorlinks   = true,    
  urlcolor     = nb,    
  linkcolor    = nb,    
  citecolor    = nb      
}

\begin{document}

\title{Engaging with AI: How Interface Design Shapes Human-AI Collaboration in High-Stakes Decision-Making}

\newcommand{\revision}[1]{{\color{black} #1}}


\author{
  \parbox[t]{0.45\textwidth}{
    \centering
    Zichen Chen \\
    \textmd{
    University of California, Santa Barbara \\
    \texttt{zichen\_chen@ucsb.edu}}
    \thanks{\textit{Corresponding author}}
  }
  \And
  \parbox[t]{0.45\textwidth}{
    \centering
    Yunhao Luo \\
    \textmd{University of California, Santa Barbara  \\
    \texttt{yunhaoluo@ucsb.edu}}
  }
  \And
  \parbox[t]{0.45\textwidth}{
    \centering
    Misha Sra \\
    \textmd{University of California, Santa Barbara  \\
    \texttt{sra@ucsb.edu}}
  }
}

\maketitle

\begin{abstract}
As reliance on AI systems for decision-making grows, it becomes critical to ensure that human users can appropriately balance trust in AI suggestions with their own judgment, especially in high-stakes domains like healthcare. However, human + AI teams have been shown to perform worse than AI alone, with evidence indicating automation bias as the reason for poorer performance, particularly because humans tend to follow AI's recommendations even when they are incorrect. In many existing human + AI systems, decision-making support is typically provided in the form of text explanations (XAI) to help users understand the AI's reasoning. Since human decision-making often relies on System 1 thinking (fast, intuitive, heuristics driven, prone to cognitive biases), users may ignore or insufficiently engage with the explanations, leading to poor decision-making. Previous research suggests that there is need for new approaches that encourage users to engage with the explanations and one proposed method is the use of  cognitive forcing functions (CFFs). In this work, we examine how various decision-support mechanisms impact user engagement, trust, and human-AI collaborative task performance in a diabetes management decision-making scenario. In a controlled experiment with 108 participants, we evaluated the effects of six distinct decision-support mechanisms split into two categories of explanations (text, visual) and four CFFs: (1) text explanations (baseline) (2) visual explanations, (3) AI confidence levels (CLs), (4) human feedback, (5) AI-driven questions, and (6) performance visualization. Our findings reveal that mechanisms such as AI CLs, text explanations, and performance visualizations significantly enhanced human-AI collaborative task performance, as measured by decision accuracy, and improved trust when AI reasoning clues were provided. While mechanisms like human feedback and AI-driven questions encouraged deeper reflection, they may have resulted in decreased task performance, likely due to increased cognitive effort and heightened scrutiny, which negatively impacted trust. Simple mechanisms like visual explanations did not significantly improve trust, highlighting the need for a contextual and balanced approach between CFF and XAI design with interactivity, decision frequency, and task complexity.


\end{abstract}

\section{Introduction}\label{intro}

\begin{figure}[t]
    \centering
    \includegraphics[width=0.9\linewidth]{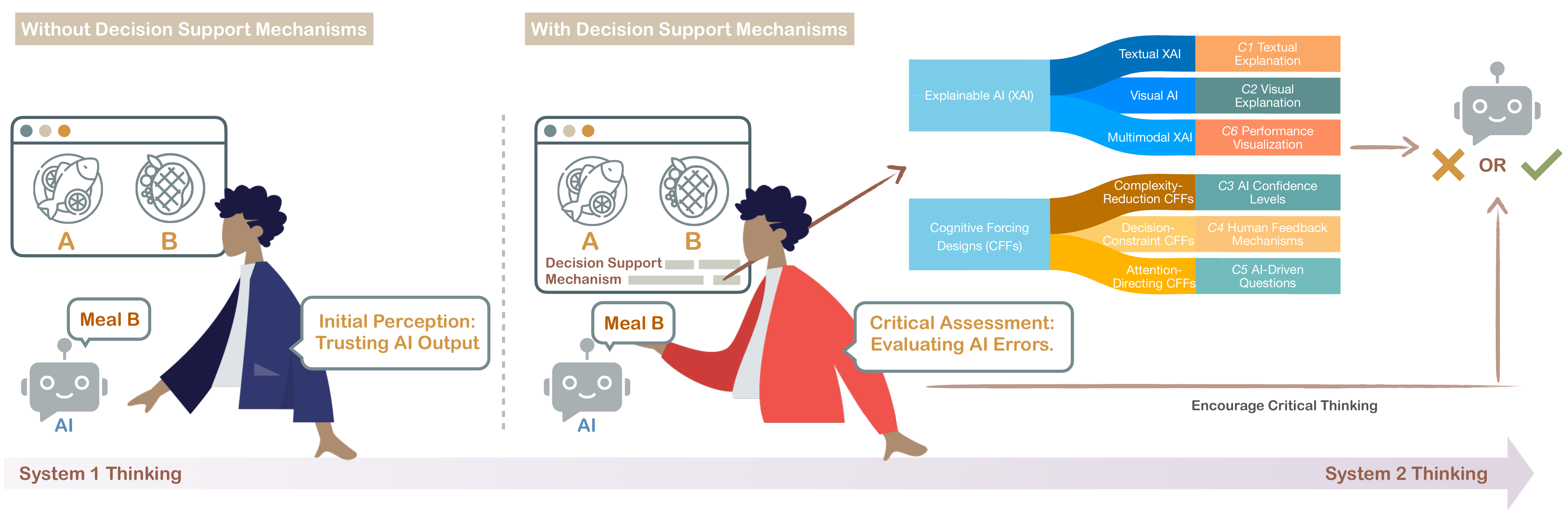}
    \caption{Dual-layer framework guiding the transition from System 1 (intuitive thinking) to System 2 (deliberative thinking) with decision support mechanisms.
    We study six decision support mechanisms, aiming to explore how explanations of varying cognitive load influence user engagement. These mechanisms are designed to encourage a transition from System 1 to System 2 processing. These mechanisms include: C1 - Textual Explanation, C2 - Visual Explanation, C3 - AI Confidence Levels (CLs), C4 - Human Feedback, C5 - AI-Driven Questions, and C6 - Performance Visualization. The framework categorizes mechanisms under XAI and CFFs.}
    \label{fig:overall}
\end{figure}

Chronic diseases like diabetes require constant informed decision making to maintain optimal health outcomes~\cite{zhao2022elements,riegel2021characteristics,ellahham2020artificial,srikanth2020type,grady2014self}. This ongoing responsibility traditionally burdens both patients and healthcare providers, requiring significant effort and attention. The emergence of artificial intelligence (AI) systems offers promising new tools to support these frequent and critical decisions~\cite{lutsker2024glucose,kraus2024data,davis2023ai,fazakis2021machine}. In particular, generative AI shows potential to help people with diabetes in meal planning to better control blood sugar levels~\cite{li2024integrated,ceballos2024open,gopalakrishnan2024recommendations,dao2024llm,belyaeva2023multimodal}. The effectiveness of human-AI collaboration in this context critically depends on users finding the right balance between trusting AI suggestions and exercising their own judgment.

Previous research has shown that trust plays a critical role in human-AI interactions, particularly in healthcare settings~\cite{zuchowski2024trust,choung2023trust,quinn2021trust}. Users may over-rely or under-trust an AI system. Over-trust can lead to unquestioned acceptance of AI suggestions, even wrong ones, potentially resulting in incorrect or harmful recommendations~\cite{shekar2024people,zhang2020effect}. Conversely, under-trust may cause users to disregard valuable AI-generated advice that could potentially improve health outcomes~\cite{lu2021human}. An appropriate balance between trust and skepticism is crucial for the safe and effective use of AI~\cite{habbal2024artificial,zerilli2022transparency,shneiderman2020bridging}, especially in healthcare where decisions can have immediate and significant consequences~\cite{nasarian2024designing,asan2020artificial}.

Recent research has explored various strategies for trust calibration in AI systems~\cite{ma2023should,buccinca2021trust,bansal2021does,li2024overconfident,zhang2020effect}. However, the challenges of overtrust and undertrust persist, preventing the ideal scenario in which human-AI collaboration outperforms humans or AI systems alone~\cite{buccinca2021trust,gajos2022people}. Even with the growing emphasis on explainable AI (XAI)~\cite{ali2023explainable,dwivedi2023explainable}, providing explanations alone has not consistently mitigated trust issues~\cite{reingold2024dissenting}. Studies suggest that overreliance persists when users are faced with complex tasks or explanations~\cite{10.1145/3613904.3642474,10.1145/3555572,logg2019algorithm}. This phenomenon can be understood through the lens of the dual process theory, which distinguishes between System 1 (fast, intuitive) and System 2 (slow, deliberative) thinking \cite{kahneman2011thinking,eisbach2023optimizing,liao2022designing,liao2021human,gajos2022people}. Users often weigh the cognitive cost (System 2) of evaluating AI suggestions against perceived benefits, potentially defaulting to uncritical acceptance (System 1) if the effort seems too high~\cite{vasconcelos2023explanations,10.1145/3610219,liao2022designing}. Conversely, under-trust can lead to the dismissal of valuable AI support, limiting the potential benefits of human-AI collaboration. Users may rely excessively on their intuitions (System 1) or engage in overly critical analysis (System 2) that can both discount AI contributions. Strategies to promote more analytical thinking and mitigate cognitive biases inherent in System 1 thinking have recently been explored \cite{buccinca2021trust}. Among these, cognitive forcing functions (CFFs) are strategically designed interventions that can interrupt automatic, intuitive reasoning at the precise moment of decision making, causing individuals to shift toward a more deliberate, analytical thought process or System 2 thinking \cite{lambe2016dual}. 

To encourage a shift from System 1 to System 2 thinking, with the goal to help overcome issues of over-trust or under-trust by enhancing engaged task performance, we designed six decision-support mechanisms including a baseline text explanation method. The six design mechanisms are grouped into two main areas: XAI (Explainable AI) and CFFs (Cognitive Framework Features). XAI focuses on enhancing the transparency of AI systems and is further divided into three categories based on their presentation modality: (1) Textual XAI, (2) Visual XAI, and (3) Multimodal XAI.
%
CFFs are designed to influence user decision-making by reducing cognitive biases and encouraging deliberate, thoughtful processes. These features are classified into three areas based on the dimension of cognitive effort:
\begin{itemize} [noitemsep, topsep=0pt]

    \item Attention-Directing CFFs: Guide user focus and prompt critical thinking (e.g., AI-driven questions).
    \item Complexity-Reduction CFFs: Reduce cognitive load by making AI outputs easier to interpret (e.g., AI confidence levels (CLs)  that transparently communicate the model's uncertainty or performance visualizations that show AI task performance over time).
    \item Decision-Constraint CFFs: Allow users to intervene in decision-making processes (e.g., human feedback mechanisms that enable validation of AI outputs).
\end{itemize}


To evaluate the effectiveness of our designed mechanisms, we conducted a human subject study involving 108 participants, with 18 participants assigned to each of the six conditions, in a diabetes meal planning scenario. Each participant was went through 3 phases in order for each of the 20 questions they answered: decision-making without AI assistance (P1), decision-making with AI assistance but without any explanation (P2), and decision-making with AI assistance accompanied by one of the six decision-support mechanisms (P3). 
Our evaluation integrates both quantitative and qualitative methods. We collected both objective and subjective data on decision accuracy, satisfaction, system complexity, reliability, trust, and perceived accuracy using pre- and post-questionnaires.  In addition, we examine user responses across dimensions such as confidence and understanding, consistency and cognitive load, ease and usefulness, trust and comfort, and trust dynamics across the designed mechanisms. Statistical analyzes are performed to assess the impact of each decision support mechanism while controlling for individual differences, including AI familiarity and educational background.

The contribution of our work is threefold. First, we provide empirical evidence on the effectiveness of different decision support mechanisms in influencing user trust, engagement, and decision-making. 
Second, we contribute to understanding of trust dynamics in human-AI interaction by exploring how different explanation types and CFFs influence trust calibration.
And lastly, we present a methodological approach for evaluating decision support mechanisms through controlled experiments, isolating the effects of individual explanation and CFF types. Our approach can be adapted to other domains beyond healthcare.

\section{Results}\label{sec:results}
Our study explores three key hypotheses regarding these mechanisms: 
\begin{enumerate}[noitemsep]
    \item \textbf{Hypothesis 1 (H1):} People tend to trust AI's answers, even when explanations are minimal, and this can lead to overreliance.
    \item \textbf{Hypothesis 2 (H2):} Access to explanation mechanisms makes users more critical of AI suggestions, encouraging deeper evaluation of AI outputs.
    \item \textbf{Hypothesis 3 (H3):} Explanation mechanisms with a balanced Explanation Information Load (EIL) that provide interpretable model reasoning support for effective human-AI collaboration by enhancing decision-making and facilitating trust calibration without overwhelming users.
\end{enumerate}

\subsection{Users Over-Rely on AI Suggestions, Even When Incorrect (H1)}

\begin{figure}[ht]
    \centering
    \begin{minipage}[c]{0.5\textwidth}
        \centering
        \resizebox{\textwidth}{!}{%
        \begin{tabular}{lccc}
        \toprule
        \textbf{AI is Incorrect} & \textbf{Correct (\%)} & \textbf{Incorrect (\%)} & \textbf{p-value} \\ 
        \midrule
        Correct & 109 (22.4\%) & 94 (19.3\%) & < 0.001 \\ 
        Incorrect & 19 (3.9\%) & 264 (54.3\%) & \\ 
        \midrule
        \textbf{AI is Correct} & \textbf{Correct (\%)} & \textbf{Incorrect (\%)} & \textbf{p-value} \\ 
        \midrule
        Correct & 1131 (67.6\%) & 62 (3.7\%) & < 0.001 \\ 
        Incorrect & 200 (11.9\%) & 281 (16.8\%) & \\ 
        \bottomrule 
        \end{tabular}
        }
        \captionof{table}{McNemar Test Results for User Decisions with AI Suggestions.}
        \label{tab:mcnemar}
    \end{minipage}%
    \hfill
    \begin{minipage}[c]{0.45\textwidth}
        \centering
        \includegraphics[width=\linewidth]{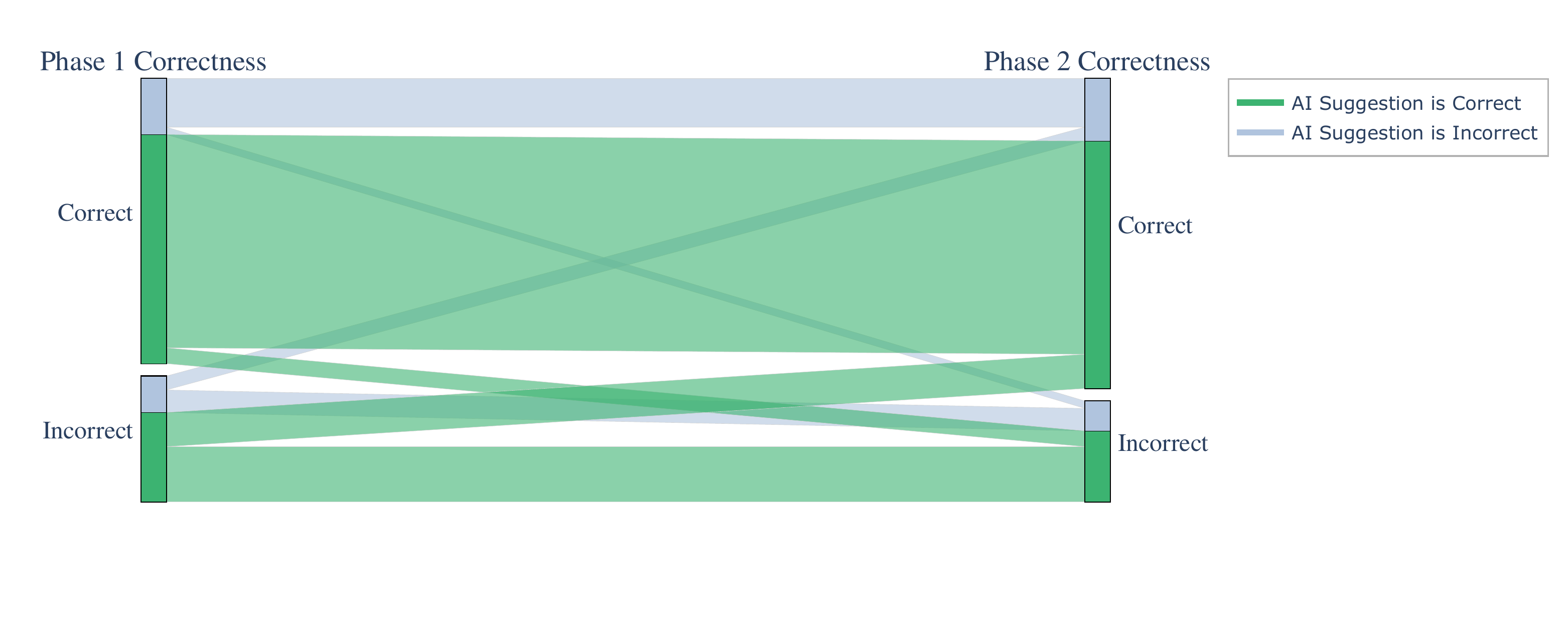}
        \caption{Visualization of User Answer Correctness.}
        \label{fig:flow-chart}
    \end{minipage}
    \end{figure}

To test the first hypothesis (H1), which posits that users tend to trust AI's suggestions, we conduct a statistical analysis exploring how users behave when the AI suggestions are both correct and incorrect. 
Specifically, we investigate whether users adjust their decisions based on the accuracy of AI's output. We chose the McNemar test for analysis, since it allows us to examine changes in user behavior before and after AI's suggestion, particularly focusing on paired nominal data. We show how user decisions shift in Table.~\ref{tab:mcnemar}: (a) when AI predictions are correct, (b) when AI predictions are incorrect. 
The test evaluates whether there is a significant change in decision patterns between these two phases. 

\textbf{When AI Suggestions are Incorrect.}\space\space
The highly significant p-value ($p < 0.001$) indicates that a notable proportion of users follow AI suggestions even when they are incorrect, demonstrating a bias toward trusting AI-generated advice. This highlights an over-reliance on AI, even in cases where the suggestions are faulty, something that prior work has also demonstrated \cite{bussone2015role, jacobs2021machine, lai2019human}.

\textbf{When AI Suggestions are Correct.}\space\space
The statistical analysis revealed a highly significant result ($p < 0.001$), showing that users are strongly inclined to accept AI's correct answers. This finding supports the hypothesis that users exhibit a high level of trust in AI, particularly when its suggestions are accurate.



\textbf{Behavioral Pattern Analysis.} \space\space
These results provide evidence supporting our hypothesis. The results suggest that people tend to follow AI suggestions with limited critical evaluation, often accepting them without thoroughly processing the outcomes. Users show a tendency to rely on AI suggestions, even when the AI is wrong. Such over-reliance can be explained by cognitive bias, where users may perceive the AI system as an authoritative figure, especially in tasks requiring complex decision-making.

When AI is correct, users demonstrate a greater degree of trust, suggesting that accurate AI assistance can reinforce user confidence in the system. However, when AI is incorrect, the lack of significant correction by the users indicates potential risks in over-reliance. This can be harmful in high-stakes applications such as healthcare decision making. We further visualize these findings in Figure.~\ref{fig:flow-chart}.


\subsection{AI Explanation Mechanisms Can Help People Engage (H2)}

We hypothesize that integrating decision support mechanisms during the decision-making process would lead to an increase in user engagement (H2). We compare engagement changes between phase 1 (P1: decision-making w/o AI assistance) to phase 2 (P2: decision-making w/ AI assistance but w/o any explanation) and between phase 1 (P1: decision-making w/o AI assistance) to phase 3 (P3: decision-making w/ AI assistance accompanied by a decision-support mechanism) using the Wilcoxon signed-rank test~\cite{woolson2005wilcoxon}, a non-parametric test suitable for analyzing paired data and comparing these engagement rates. We also compare the pre- and post-questionnaire results to measure the engagement level. The questionnaires include questions related to satisfaction, system complexity, reliability, trust, and system accuracy. More details on the questionnaires and measurement can be found in Method~\ref{appdeix:measures}.

\textbf{Condition 1: Textual Explanations.}\space\space
Textual explanations (C1), as our baseline condition, includes detailed textual explanations of the AI's decision-making process. There is a significant increase in engagement ($p=0.049$), indicating that textual explanations help users to understand AI decisions through a logical structure. By comparing the pre- and post-questionnaire results, we find a significant improvement in system reliability ($p=0.048$), suggesting that such explanations can improve user's trust in the AI system. However, it is important to note that the system complexity also increased ($p=0.008$), which may have led to increased cognitive overload.

\textbf{Condition 2: Visual Explanations.}\space\space 
Visual explanations (C2) include a basic visual explanation of the AI's attention without any additional feedback mechanism. The results do not show any significant engagement improvement ($p=0.776$), suggesting that visual explanations alone may not be sufficient to encourage deeper engagement. Users likely accepted the AI's output without reflecting on its reasoning, which points to potential over-reliance.

\textbf{Condition 3: AI Confidence Levels.}\space\space
AI CL (C3) allows users to view the AI's and their own confidence levels (CLs). The results show a significant increase in engagement ($p=0.001$), indicating that confidence levels may help users to calibrate their trust in the AI system. CL is a form of transparency that provides users with direct insight into the AI's certainty, which encourages users to critically compare their judgment with the AI's. Users are not only passive recipients of AI suggestions but active participants in the decision-making loop, which may explain the increased engagement. However, we observe that there is no significant improvement in system reliability and trust, suggesting that confidence levels alone may not be sufficient to build trust.

\textbf{Condition 4: Human Feedback.}\space\space
Similar to AI CLs (C3), human feedback (C4) allows users to input their CLs. It shows a significant increase in engagement ($p=0.006$), suggesting that allowing users to input their own thinking encourages introspection and deep cognitive involvement in decision-making. However, same to AI CLs (C3), there is no significant improvement in system reliability and trust. This indicates that while user input can increase engagement, it may not necessarily modulate trust in the AI system.

\textbf{Condition 5: AI-Driven Questions.}\space\space
AI-driven questions (C5) include AI-generated questions to prompt users to reflect on AI's suggestions. The results show a borderline significance ($p=0.058$), indicating that while AI-generated reflective questions encourage some degree of engagement, they are not as effective as other conditions. The questions require users to think deeper about the AI's decision, which forces user's to shift from System 1 to System 2 thinking. The borderline result suggests that AI-driven questions may  hold promise, if optimized.

\textbf{Condition 6: Performance Visualization.}\space\space
Performance visualization (C6) includes a visualization of the AI and user's task performance over time. It shows a significant increase in engagement ($p=0.009$), suggesting that the visualized comparison of human and AI performance contributed to a reflective decision-making process. Perhaps it allows users to contextualize AI suggestions based on empirical evidence, forcing the user to evaluate AI outputs more deliberately. The visualized comparison is intuitive and easy to process by the human, leading to higher engagement.

\subsection{Explanation Information Load and Interpretable Reasoning Enhance Human-AI Collaboration (H3)}

We calculated an Explanation Information Load (EIL) across the six conditions. The details of EIL can be found in Method~\ref{sec:EPE}. The resulting values represent the information load on the users, that needed to be cognitively processed, for each condition. 

The results in Table~\ref{tab:epe_results} show notable variation in processing effort between the conditions.

\begin{table}[ht]
\centering
\resizebox{\linewidth}{!}{
\begin{tabular}{lcccccc}
\hline
\textbf{Conditions} &
\textbf{Textual Explanations (C1)} & \textbf{Visual Explanations (C2)} & \textbf{AI CLs (C3)} & \textbf{Human Feedback (C4)} & \textbf{AI-Driven Questions (C5)} & \textbf{Performance Visualization (C6)} \\ \hline
\textbf{EIL} &
1.520 & 1.749 & 0.602 & 1.204 & 1.965 & 0.602 \\ \hline
\end{tabular}
}
\vspace{5pt}
\caption{Explanation Information Load (EIL) across six AI explanation mechanisms.}
\label{tab:epe_results}
\end{table}

Based on our hypothesis (H2), we expect to see a balanced cognitive load for better human-AI collaboration (H3). Across the six conditions, we observe a variety of outcomes, ranging from significant improvements in user accuracy to negligible or even borderline impacts. 

In visual explanations (C2) and AI CLs (C3), where AI explanation mechanisms are designed to provide transparency and interpretability, we observe a significant increase in user accuracy ($p=0.018$ and $p=0.004$). 
The moderate EIL in visual explanations (C2) ($EIL=1.749$) provides users with comprehensive but understandable insights into AI reasoning, encouraging critical engagement without overwhelming them. 
The low EIL in AI CLs (C3) ($EIL=0.602$) allows users to judge the certainty of AI suggestions at a glance, serving as an efficient cognitive aid. 
These mechanisms provide users with interpretable model reasoning that conveys the AI's rationale and confidence, respectively. This combination of a manageable EIL and interpretable reasoning support users in making informed decisions without information overload.

In contrast, mechanisms that lack clear interpretive reasoning exhibit less impact on performance. Performance visualization (C6), with an EIL of 0.60, demonstrates borderline significance ($p=0.092$), suggesting that while users are encouraged to engage with historical performance data, the lack of explicit interpretive guidance left them to draw their own conclusions from visual information. This aligns with cognitive psychology research, which indicates that high information density, without sufficient interpretive support, can lead to cognitive strain, ultimately limiting performance and reducing trust calibration~\cite{arnold2023dealing}.

Furthermore, in visual explanations (C2) ($EIL = 1.520$), human feedback (C4) ($EIL = 1.204$), and AI-driven questions (C5) ($EIL = 1.965$), we found no significant improvement in user performance ($p=0.673$, $p=0.674$, and $p=0.302$). These conditions either lack a direct interpretive model component or introduce high cognitive demands without adequate support. For example, textual explanations (C1) present intuitive but limited information, which may not provide users with the depth needed to make well-calibrated decisions. Human feedback (C4) encourages self-assessment, but without specific reasoning cues from the AI, it does not significantly enhance decision accuracy. AI-driven questions (C5), with the highest EIL, likely lead to information overload, as users are required to reflect deeply on their decisions.

These results suggest that EIL and the presence of interpretable model reasoning are important for effective human-AI collaboration. 
Mechanisms that balance a manageable EIL with clear reasoning cues can improve collaboration performance. In contrast, mechanisms that lack interpretive support or impose high cognitive demands may hinder performance, highlighting the importance of designing AI explanation mechanisms that offer clear interpretive insights aligned with the user's cognitive capacity.

\begin{table}[ht]
\centering
\begin{tabular}{l|c@{\hspace{1em}}c|c@{\hspace{1em}}c|l}
\hline
& \multicolumn{2}{c|}{Mean} & \multicolumn{2}{c|}{Standard Deviation} & \multirow{2}{*}{Significance} \\
\cline{2-5}
& Before & After & Before & After & \\
\hline
\multicolumn{6}{l}{\textbf{Textual Explanations (C1)}} \\
\hline
Satisfaction & 5.222 & 5.556 & 0.786 & 1.165 & $p = 0.122$ \\
System Complexity & 5.111 & 5.778 & 0.936 & 0.916 & $p = 0.008^*$ \\
Reliability & 4.722 & 5.556 & 1.239 & 1.343 & $p = 0.048^*$ \\
Trust & 5.167 & 4.889 & 1.167 & 1.242 & $p = 0.368$ \\
Perceived Accuracy & 5.111 & 5.444 & 0.809 & 1.066 & $p = 0.272$ \\
\hline
\multicolumn{6}{l}{\textbf{Visual Explanations (C2)}} \\
\hline
Satisfaction & 5.278 & 5.778 & 0.931 & 0.629 & $p = 0.050^*$ \\
System Complexity & 5.944 & 6.222 & 0.911 & 0.786 & $p = 0.374$ \\
Reliability & 5.278 & 6.000 & 0.989 & 0.667 & $p = 0.005^*$ \\
Trust & 5.333 & 5.056 & 1.106 & 0.911 & $p = 0.420$ \\
Perceived Accuracy & 5.333 & 5.611 & 0.745 & 0.826 & $p = 0.166$ \\
\hline
\multicolumn{6}{l}{\textbf{AI CLs (C3)}} \\
\hline
Satisfaction & 5.333 & 5.722 & 1.106 & 0.731 & $p = 0.244$ \\
System Complexity & 5.833 & 6.111 & 0.833 & 1.197 & $p = 0.407$ \\
Reliability & 5.000 & 5.722 & 1.414 & 0.803 & $p = 0.114$ \\
Trust & 5.222 & 5.000 & 0.853 & 0.745 & $p = 0.206$ \\
Perceived Accuracy & 5.000 & 5.667 & 1.291 & 0.745 & $p = 0.039^*$ \\
\hline
\multicolumn{6}{l}{\textbf{Human Feedback (C4)}} \\
\hline
Satisfaction & 4.889 & 5.222 & 1.197 & 1.181 & $p = 0.143$ \\
System Complexity & 5.444 & 5.667 & 1.212 & 1.247 & $p = 0.310$ \\
Reliability & 4.500 & 4.778 & 1.213 & 1.474 & $p = 0.318$ \\
Trust & 4.611 & 4.278 & 1.208 & 1.592 & $p = 0.417$ \\
Perceived Accuracy & 4.722 & 4.778 & 1.239 & 1.315 & $p = 0.592$ \\
\hline
\multicolumn{6}{l}{\textbf{AI-Driven Questions (C5)}} \\
\hline
Satisfaction & 5.556 & 5.611 & 0.896 & 1.113 & $p = 0.891$ \\
System Complexity & 5.889 & 5.500 & 0.657 & 1.344 & $p = 0.218$ \\
Reliability & 5.444 & 5.444 & 1.066 & 1.165 & $p = 0.857$ \\
Trust & 5.667 & 5.278 & 0.667 & 0.931 & $p = 0.100$ \\
Perceived Accuracy & 5.611 & 5.611 & 0.891 & 0.891 & $p = 0.951$ \\
\hline
\multicolumn{6}{l}{\textbf{Performance Visualization (C6)}} \\
\hline
Satisfaction & 5.000 & 4.944 & 1.202 & 1.471 & $p = 0.773$ \\
System Complexity & 5.333 & 4.056 & 1.291 & 1.545 & $p = 0.008^*$ \\
Reliability & 4.778 & 4.333 & 1.436 & 1.528 & $p = 0.210$ \\
Trust & 5.056 & 4.278 & 1.224 & 1.626 & $p = 0.032^*$ \\
Perceived Accuracy & 4.722 & 4.444 & 1.283 & 1.771 & $p = 0.543$ \\
\hline
\end{tabular}
\vspace{5pt}
\caption{Within-Condition statistical results for satisfaction, system complexity, reliability, trust, and perceived accuracy across all conditions: Textual Explanations (C1), Visual Explanations (C2), AI CLs (C3), Human Feedback (C4), AI-Driven Questions (C5) and Performance Visualization (C6). }
\label{tab:within_results}
\end{table}

\subsection{Textual Explanations (C1): The Impact of Basic AI Support on 
Decision-Making}

In textual explanations (C1), participants interact with the AI system that provides support but without any additional explanatory mechanisms, feedback loops, or engagement-enhancing features. We evaluate the impact of this condition's effectiveness with respect to user satisfaction, system complexity, reliability, trust, and perceived accuracy. We also apply these metrics to other conditions. Detailed statistical results for all conditions can be found in Table~\ref{tab:within_results}.

\textbf{Satisfaction.}\space\space
While there is a modest increase in satisfaction metrics, the improvement is not statistically significant ($p=0.122$). The results show that while textual explanations may provide some additional clarity, they do not have a substantial impact on users' overall satisfaction with the system. The relatively high variability in post-satisfaction ($Std=1.17$) indicates that user responses are mixed, with some satisfied with the explanations while others may feel they add little value.

\textbf{System Complexity.}\space\space
We observe a significant increase in system complexity ($p=0.008$) after users interact with textual explanations (C1). The result suggests that textual explanations may introduce additional complexity burden on users, making the system more challenging to use. Although textual explanations are designed to clarify the AI's decision-making process, they can also make the system complex, especially if the explanations are not easily digestible or require additional effort to interpret.

\textbf{Reliability.}\space\space
There is a statistically significant increase in reliability ($p=0.048$) in textual explanations (C1). It indicates that textual explanations have a positive impact on how users perceive the AI system's reliability. Textual explanation as a form of transparency helps users better understand the rationale behind AI decisions, building users' confidence in the AI's outputs. The results suggest that even though explanations may increase system complexity, they also enhance user perception of the system's reliability. It shows that explanations can calibrate user trust on AI systems and help them assess AI suggestions more critically.

\textbf{Trust.}\space\space
Interestingly, user trust in the AI system slightly decreased, though this change was not statistically significant ($p=0.368$). 
Despite the increase in the reliability, the additional complexity of textual explanations may have increased cognitive load, leading to a minor reduction in overall trust. The results highlight a key challenge in AI design: while explanations can increase reliability, they should be presented in a way that does not increase cognitive processing load to the point where trust is negatively affected. We can also find there is a mismatch between the user's mental model of the AI and the actual usefulness of the explanations provided. If explanations do not align well with user mental models, they may lead to over-reliance or skepticism toward the AI system.

\textbf{Perceived Accuracy.}\space\space
Perceived accuracy shows a modest increase, but without statistical significance ($p=0.272$). The results suggest that while textual explanations may help users better understand the AI's decision-making process, they do not universally improve user perceptions of the AI's accuracy. The variability in post-study accuracy ratings ($Std=1.07$) indicates a wide range of user experiences, with some participants finding the explanations helpful to assess AI accuracy, while others do not.

\subsection{Visual Explanations (C2): The Impact of Text Explanations on Decision-Making}
In visual explanations (C2), participants interact with an AI system that enhances transparency by highlighting key reasoning areas within the visual input and labeling them, directly mapping the AI’s decision-making process and the visual evidence.  Text explanations are widely regarded as a means of improving AI interpretability~\cite{joyce2023explainable,chen2023xplainllm}, while visual explanations offer an intuitive way for users to verify and contextualize the AI's reasoning~\cite{Kim2022HIVE}.

\textbf{Satisfaction.}\space\space
We observe a statistically significant increase in satisfaction ($p=0.050$), before and after interacting with visual explanations (C2), suggesting that participants feel more content with the system after using it. The results highlight the potential for even simple mechanisms to improve the user experience in decision-making tasks. Though the improvement in satisfaction is modest, it shows that users are receptive to explanation mechanisms, as these mechanisms can help users navigate complex decision-making tasks more effectively. It is also worth noting that this improvement may plateau without further engagement mechanisms, as evidenced by the results for trust and engagement discussed below.

\textbf{System Complexity.}\space\space
System complexity remains consistently low, with no significant changes before and after interacting with visual explanations (C2). The result implies that the visual explanation mechanism does not introduce additional complexity or cognitive burden on users. Participants generally find the system easy to use. However, while simplicity is desirable, a lack of change may also indicate users are not engaging deeply with the AI system. 

\textbf{Reliability.}\space\space
We observe a statistically significant increase in perceived reliability ($p=0.005$), demonstrating that users feel the AI system is more reliable after they are shown the visual explanations (C2). The result suggests that even without detailed textual explanations, the AI system's performance in supporting decisions can enhance its perception of reliability. The results suggest that users may find the system to be reliable when the AI system provides reasonably accurate suggestions, even without a clear understanding of its decision-making process. However, this gain in reliability does not translate into a corresponding increase in trust, as shown in the next subsection, pointing toward potential limitations in how trust is developed in AI systems without transparency or interpretability.

\textbf{Trust.}\space\space
While reliability improves significantly, there is no evidence of a corresponding increase in trust ($p=0.421$). The results suggest that users may perceive the AI system as reliable based on its performance, but they do not trust it to guide their decisions under the current conditions. Trust in AI systems depends not only on how reliable the system appears but also on how transparent and interpretable the system is, and how much agency the user has. Without explanations or feedback, users may begin to feel uncertain about how to critically assess its outputs, leading to a trust decline.

\textbf{Perceived Accuracy.}\space\space
Perceived accuracy shows a small, non-significant increase in visual explanations (C2). While participants improve their views on the accuracy of the AI's suggestions, the improvement does not show statistical significance. 
The results suggest that while the AI is generally perceived as accurate, the lack of system transparency may limit user ability to engage critically with the AI's outputs with impact on perception of credibility and trust.

\subsection{AI CLs (C3): The Impact of CLs on Decision-Making}
In AI CLs (C3), participants receive AI's CLs for each suggestion, providing a quantifiable metric of the AI's certainty. CL is a form of transparency that provides users with a clear sense of the system's certainty in its outputs. We also provide a CL to estimate user performance to help participants compare their behavior with the AI's. Different from textual explanations, which provide detailed logical reasoning, 
CLs offer explicit and quantitative measures of AI confidence.

\textbf{Satisfaction.}\space\space
While there is a small increase in satisfaction, the change is not statistically significant. The results suggest that although CLs may provide additional clarity to users, they do not greatly enhance the overall satisfaction with the system. The relatively stable satisfaction levels across pre- and post-evaluations imply that CLs provide supplementary rather than transformative information to the user and, therefore, have minimal impact on satisfaction.

\textbf{System Complexity.}\space\space
There is no significant change in system complexity after users interact with AI CLs (C3). The results suggest that while the the confidence levels introduce additional information, it is not overwhelming. The relatively high initial complexity ratings indicate that participants already perceive the AI system as moderately complex, and the addition of CLs do not dramatically change this perception.
The results suggest that CLs may not require deep cognitive effort to interpret. 

\textbf{Reliability.}\space\space
The results show a clear increase in reliability, though this change does not reach statistical significance. It suggests that CLs help users feel more confident in the AI system's reliability. While not statistically significant, the improvement in reliability points to the potential of CLs as a tool for trust calibration, enabling users to compare their performance with the AI's and adjust their trust levels accordingly.

\textbf{Trust.}\space\space
Although trust in the AI slightly decreased after the introduction of CLs, this change is not statistically significant ($p=0.206$). The results indicate a subtle interaction between the reliability and trust of the system. While CLs may improve user perception of the AI's reliability, it is possible that users become more critical of the AI suggestions when they can see that the system is not always highly confident or more accurate than a human.

\textbf{Perceived Accuracy.}\space\space
There is a significant increase in perceived accuracy after users interact with AI CLs (C3) ($p=0.039$). The result highlights the value of transparency in improving user perception of AI performance. By comparing the CLs of their performance with AI's, it helps users make more informed decisions and adjust their reliance on the system. The significant increase also suggests that CLs play an important role in calibrating trust, enabling users to align their expectations with the AI's internal assessments of its own performance.

\subsection{Human Feedback (C4): The Impact of Human Feedback on Decision-Making}

In human feedback (C4), participants need to input both their own CL and their estimation of the AI's confidence in its suggestion. 
Human feedback (C4) is designed to investigate how user engagement with the AI system may change when they are enforced to reflect on both their own ability and the AI's. 

\textbf{Satisfaction.}\space\space
Satisfaction increased slightly after interacting with human feedback (C4), though the change is not statistically significant. The relatively stable satisfaction levels across pre- and post-evaluations imply that human feedback may not be as transformative as other explanation mechanisms in the decision-making process for the current task. The lack of significance indicates that the added cognitive task of self-reflection does not substantially enhance overall satisfaction, potentially because it may introduce additional cognitive effort.

\textbf{System Complexity.}\space\space
There is a slight and non-significant increase in perceived complexity ($p=0.310$), reflecting the additional complexity of asking users to input CLs. The results show that while the added interaction may increase the cognitive processing load, users generally find the system manageable.

\textbf{Reliability.}\space\space
Reliability ratings improve slightly, but this change is not statistically significant ($p=0.318$). The results suggest that while human feedback may help users calibrate their trust in the AI system, it does not significantly impact their perception of the system's reliability. It shows that while reflection on their own confidence and the AI's confidence may lead users to feel slightly more confident in the system's reliability, the task of inputting CLs does not produce a major shift in how users perceive the AI's reliability. One potential reason is that users may have struggled to accurately judge the AI's confidence or to compare it meaningfully with their own.

\textbf{Trust.}\space\space
Trust in the AI system slightly decreased after participants have to reflect on their own CLs and those of the AI. While this change is not statistically significant ($p=0.417$), the slight decline suggests that the process of reflecting on confidence might lead some users to become more critical of the AI's suggestions and question the system's accuracy when they think their confidence is not aligned. 
This finding highlights the potential challenge of trust calibration when users are given more control over the decision-making process without clear guidance on how to interpret CLs.

\textbf{Perceived Accuracy.}\space\space
Perceived accuracy remains largely unchanged, with a very slight and non-significant increase ($p=0.592$). 
The results suggest that allowing users to input CLs does not substantially affect their perception of how accurate the AI's decisions are. It is possible that users find it challenging to interpret their CLs and align with those of the AI, limiting the impact of this interaction on perception of the AI's accuracy.

\subsection{AI-Driven Questions (C5): The Impact of AI-Generated Questions on Decision-Making}

In AI-driven questions (C5), participants are presented with AI-generated questions designed to encourage critical thinking and deeper engagement. The questions are intended to prompt users to actively evaluate the AI's suggestions, rather than passively accepting them. 

\textbf{Satisfaction.}\space\space
We observe that satisfaction slightly decreases, although this change is not statistically significant ($p=0.773$). The results suggest that while the reflective questions may help some users engage more deeply, they do not universally enhance user  satisfaction with the system. It is possible that questions may introduce additional cognitive processing load, leading to this slight decrease. 

\textbf{System Complexity.}\space\space
Interestingly, there is a statistically significant decrease in system complexity after participants interact with AI-driven questions (C5) ($p=0.008$). 
The counterintuitive result suggests that the AI's reflective questions may help users focus their decision-making process, thereby simplifying the task. By guiding users toward key considerations, the AI-generated questions might reduce cognitive overload, offering a structured framework for evaluating the AI's recommendations. The results also highlight the potential of guided reflection as a tool to make complex decision-making processes feel more manageable.

\textbf{Reliability.}\space\space
We find a slight increase in reliability, though this change is not statistically significant ($p=0.210$). The results show that while the reflective questions may encourage users to think more critically about the AI's suggestions, they do not significantly impact user perception of the system's reliability. When users are prompted to engage more deeply with AI decisions, they may become more aware of potential errors, which can undermine their confidence in the system's reliability. From the results, we  find that the reflective questions encourage deep reflection but may simultaneously reveal errors in the AI's reasoning, thus impacting how users perceive the system's reliability.

\textbf{Trust.}\space\space
We observe a statistically significant decrease in trust after participants interact with AI-driven questions (C5) ($p=0.032$). 
The result suggests that, while reflective questions are designed to improve engagement, they may also lead users to question the AI's recommendations more critically, potentially leading to a loss of trust when the AI's logic does not align with the user's expectations.
This finding highlights a critical challenge in human-AI collaboration: while encouraging users to engage critically with AI outputs can promote better decision-making, it can also expose the limitations of the AI, leading to reduced trust. 

\textbf{Perceived Accuracy.}\space\space
Perceived accuracy decreases slightly, but this change is not statistically significant ($p=0.543$). The results suggest generated questions do not substantially impact user perception of the AI's accuracy. The variability in post-accuracy ratings ($Std=1.77$) indicates a wide range of user experiences, with some participants possibly found the reflective questions helpful, while others may have struggled with interpreting the AI's logic, leading to doubts about its accuracy.

\subsection{Performance Visualization (C6): The Impact of Performance Visualization on Decision-Making}

In performance visualization (C6), participants are presented with performance visualizations that compare the AI's historical performance on the ongoing task against their own decisions. Performance visualization (C6) is intended to provide users with a clear, empirical basis for evaluating the AI's recommendations over time, thereby enabling users to calibrate their trust more effectively.

\textbf{Satisfaction.}\space\space
The participant satisfaction with the AI system remains largely unchanged after the introduction of performance visualization. The results show that while the visualization may provide useful insights into past performance, it does not have a strong impact on the overall user experience. It indicates that users view the visualizations as supplementary rather than central to their decision-making process.

\textbf{System Complexity.}\space\space
There is a slight decrease in system complexity in performance visualization (C6), though this change is not statistically significant ($p=0.218$). 
The visualizations may provide users with a clearer understanding of the AI's decision-making performance over time, helping them evaluate the AI's consistency and accuracy without overwhelming them with additional information load. 
However, the non-significant results imply that the visualizations do not substantially change user perception of system complexity.

\textbf{Reliability.}\space\space
Interestingly, participant perception of the AI system's reliability remains unchanged. The result suggests that while the visualization provided a historical perspective on the AI's accuracy, it was not compelling enough to significantly shift user views on how reliable the system is.

\textbf{Trust.}\space\space
We observe a slight decrease in trust, though this change is not statistically significant ($p=0.099$). The results suggest that the performance visualizations (C6) lead users to question the AI's recommendations more critically, potentially leading to a loss of trust when the AI's behavior does not align with the user's expectations. When users have opportunity to compare their own past performance with AI's, they may become more critical of the AI's suggestions, particularly if they identify discrepancies between their judgments and the AI's.

\textbf{Perceived Accuracy.}\space\space
Perceived accuracy remains the same before and after interacting with performance visualization (C6). This result indicates that providing participants with a visual comparison of their performance and the AI's, does not significantly impact their perceptions of the AI's accuracy. It highlights that 
while historical performance data provides users with additional information, users may rely more on the AI's immediate output when forming judgments about accuracy.

\subsection{Cross-Condition Analysis: Confidence, Cognitive Load, Usefulness, Trust and Trust Dynamics}

\begin{figure}[t]
    \centering
    \begin{subfigure}{0.3\textwidth}
        \includegraphics[width=\linewidth]{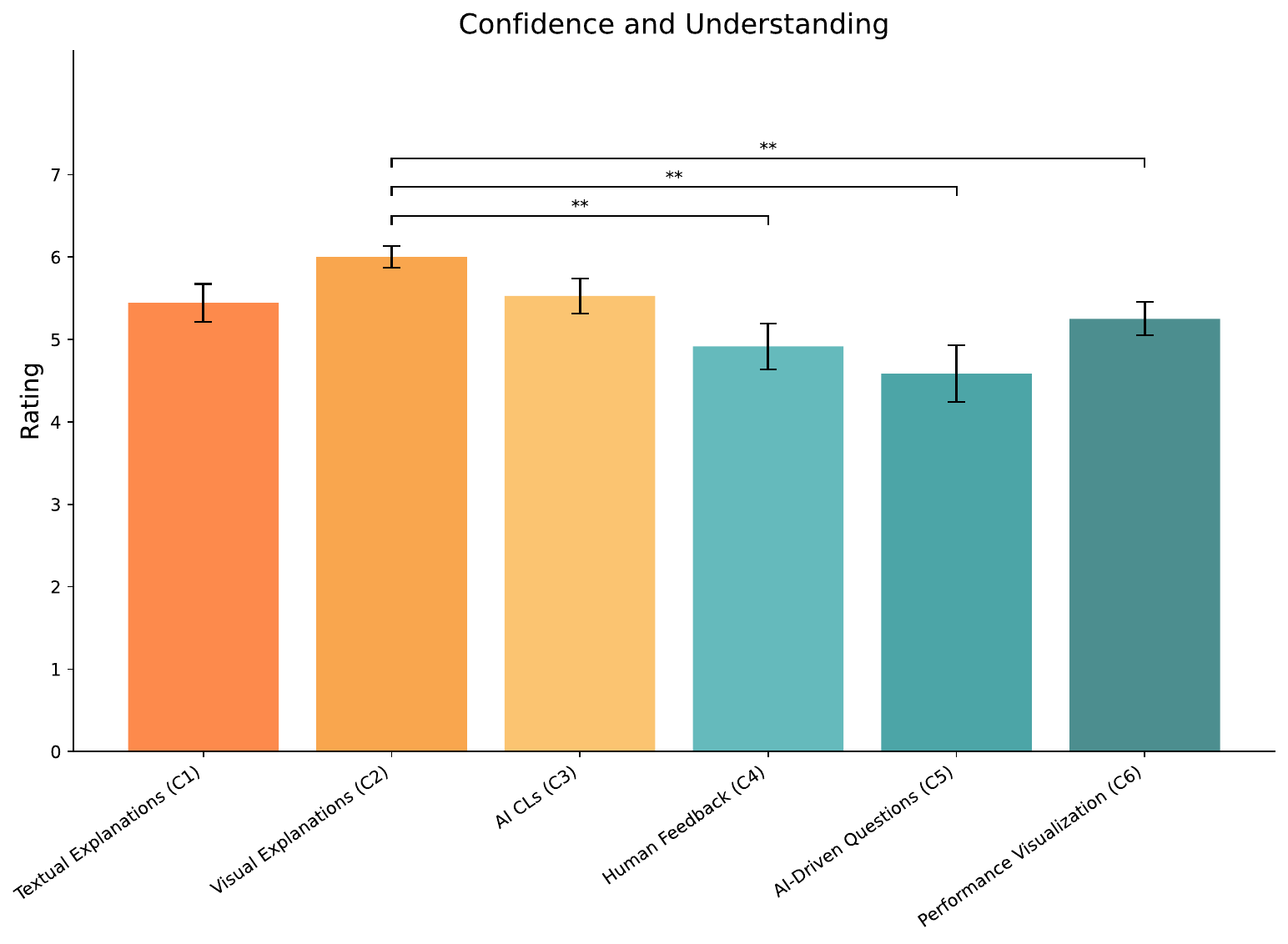}
        \label{fig:subfig1}
    \end{subfigure}
    \hfill
    \begin{subfigure}{0.3\textwidth}
        \includegraphics[width=\linewidth]{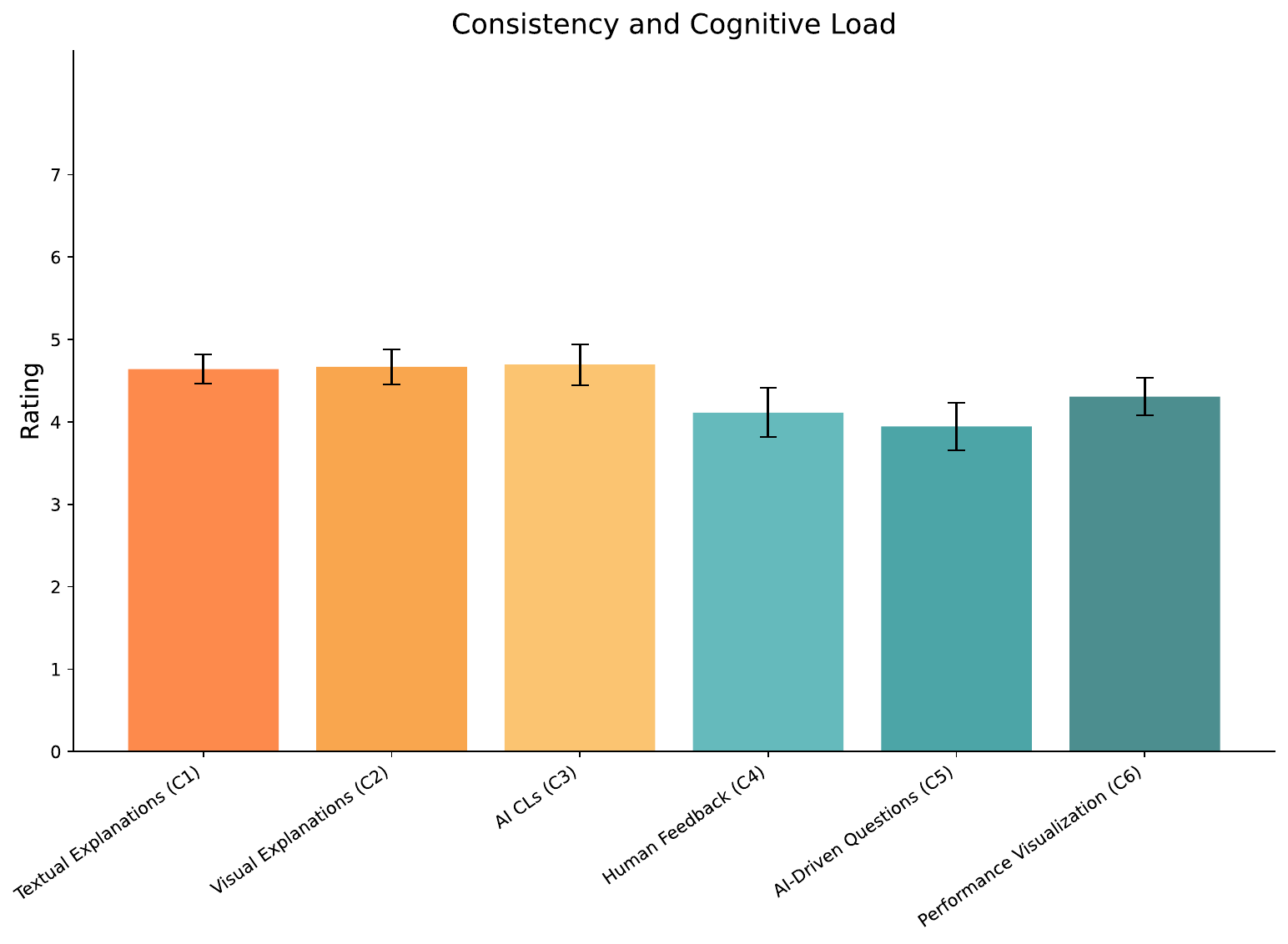}
        \label{fig:subfig2}
    \end{subfigure}
    \hfill
    \begin{subfigure}{0.3\textwidth}
        \includegraphics[width=\linewidth]{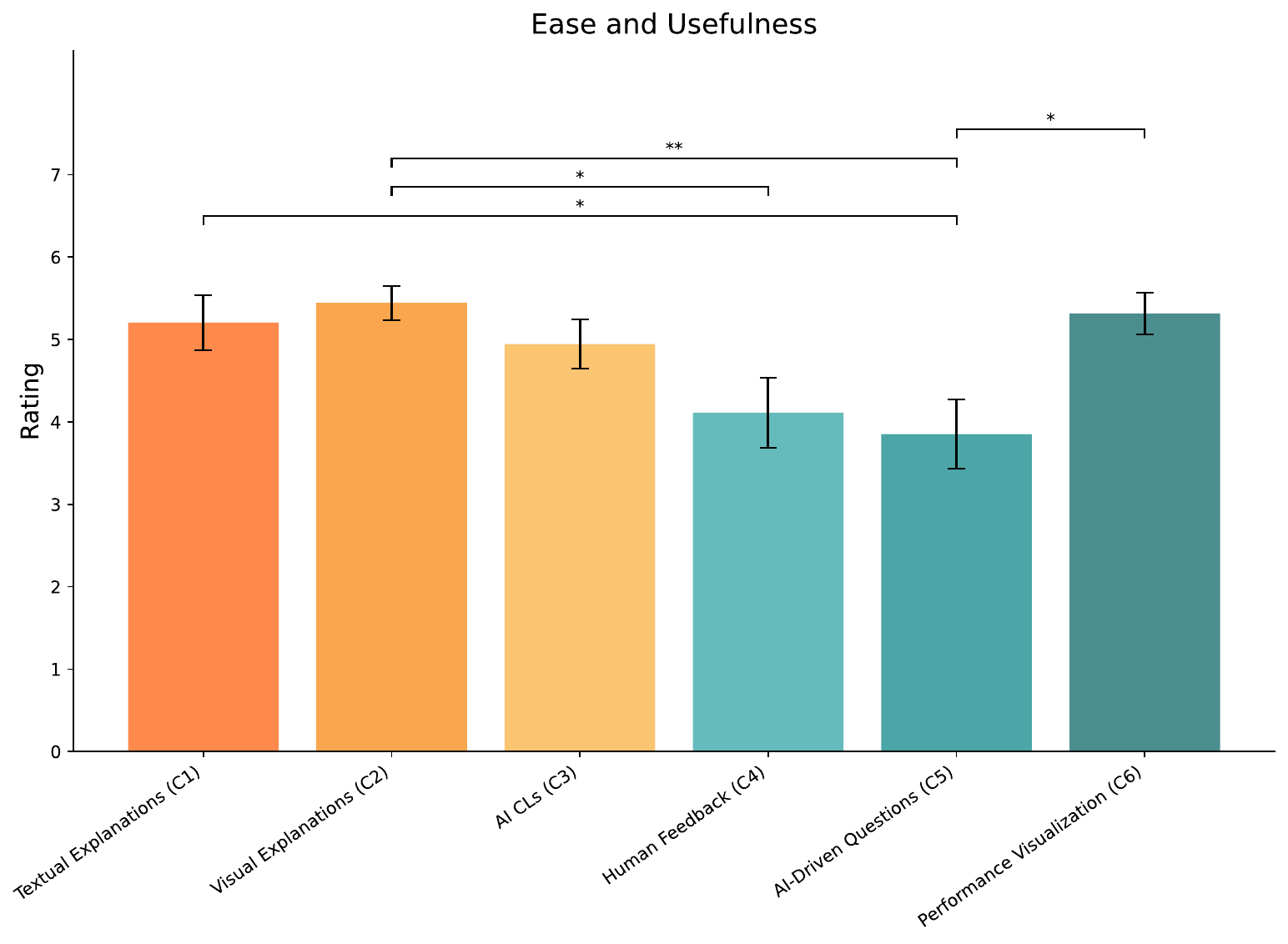}
        \label{fig:subfig3}
    \end{subfigure}
    \begin{subfigure}{0.3\textwidth}
        \includegraphics[width=\linewidth]{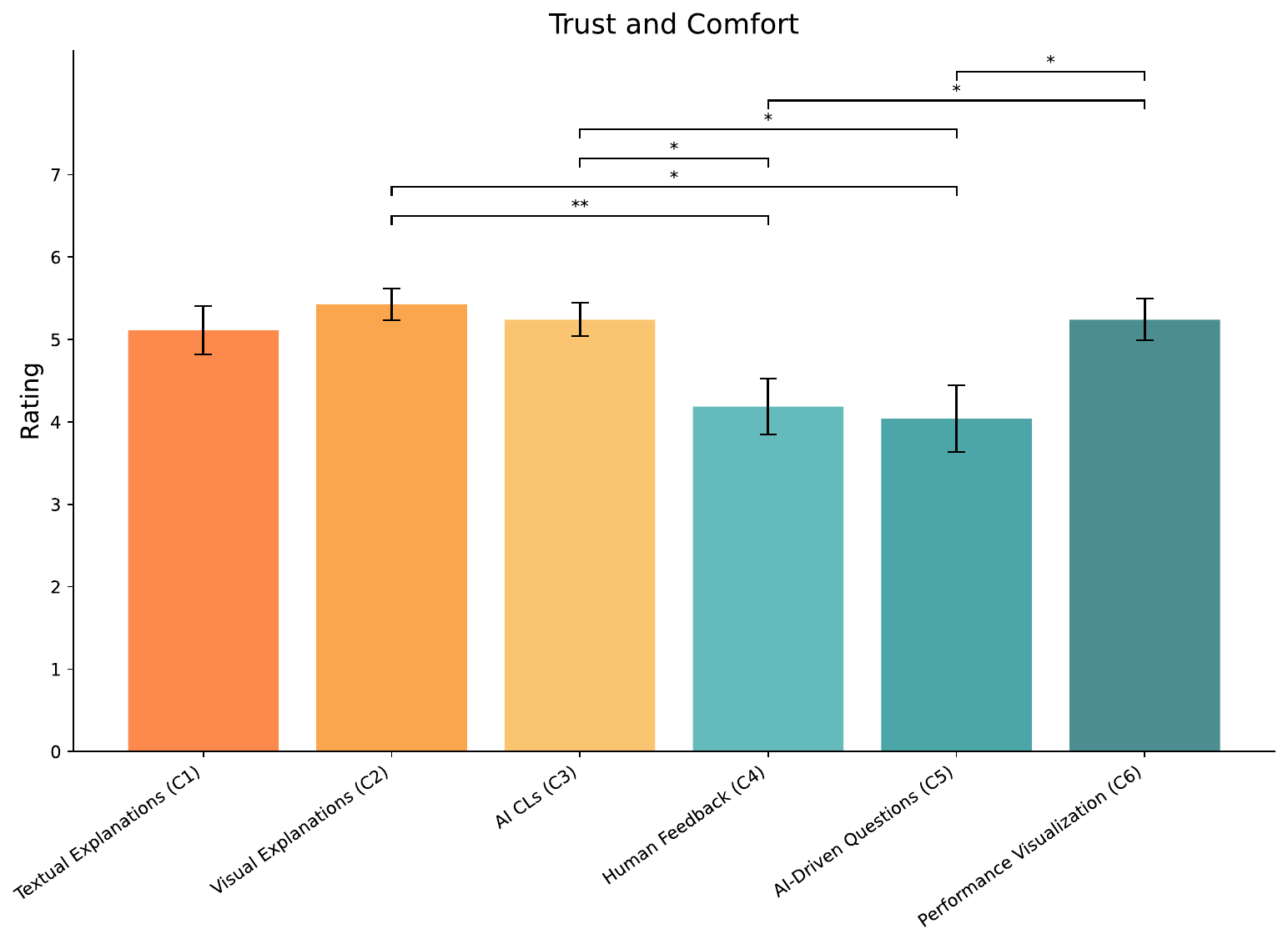}
        \label{fig:subfig4}
    \end{subfigure}
    \begin{subfigure}{0.3\textwidth}
        \includegraphics[width=\linewidth]{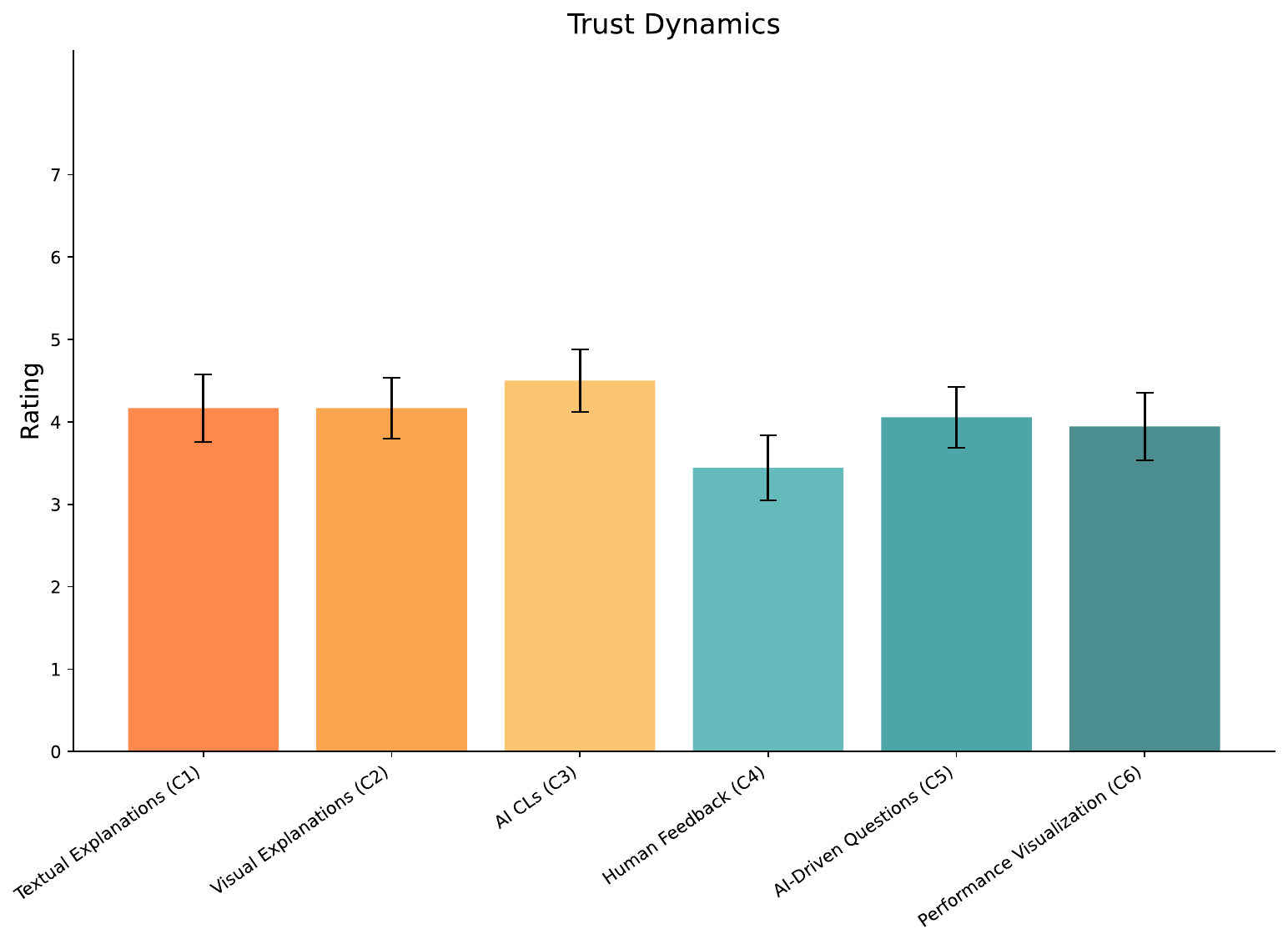}
        \label{fig:subfig5}
    \end{subfigure}
    \caption{Questionnaire scores across six conditions on the five measures: satisfaction, system complexity, reliability, trust, and system accuracy. Error bars represent variability, and horizontal bars indicate statistical significance ($p<0.05$) where applicable.}
    \label{fig:mainfig}
\end{figure}

We analyze user decision-making patterns across all the conditions to understand how different mechanisms influence key factors of human-AI collaboration. We evaluate the dimensions of \textit{confidence and understanding}, \textit{consistency and cognitive load}, \textit{ease and usefulness}, \textit{trust and comfort}, and \textit{trust dynamics}. Detailed statistical results for all conditions can be found in Figure~\ref{fig:mainfig}.

We observe that there is a significant difference in \textit{confidence and understanding} across conditions: textual explanations (C1) ($p=0.007$), human feedback (C4) ($p=0.002$), AI-driven questions (C5) ($p=0.003$), and performance visualization (C6) ($p=0.008$). Interestingly, textual explanations (C1) demonstrates the highest confidence mean ($mean=6.000$, $std=0.553$), reflecting the fact that basic mechanisms, while lacking transparency, can still evoke strong confidence, perhaps due to simplicity. In contrast, conditions 4 and 5, which require users to engage deeply, have significantly lower ratings. It suggests that the increased cognitive load leads users to question AI suggestions more critically.

The results for \textit{consistency and cognitive load} show no significant differences across conditions ($p=0.263$). Mean scores are relatively similar across all conditions, with AI CLs (C3) having a slightly higher consistency scores ($mean=4.694$, $std=1.056$) compared to AI-driven questions (C5) ($mean=3.944$, $std=1.212$). We observe that CLs may help users perceive the AI system as more consistent, because CLs provide users with transparency about the AI's certainty in its decisions. On the other hand, mechanisms that require deeper cognitive engagement (e.g., AI-driven questions (C5)) may introduce additional cognitive load without significantly improving user perception of consistency.

The \textit{ease and usefulness} dimension shows significant differences across conditions ($p=0.029$), with textual explanations (C1) outperforming human feedback (C4) ($p=0.025$) and AI-driven questions (C5) ($p=0.009$). Condition 1 shows the highest scores ($mean=5.444$, $std=0.882$), reaffirming the simple explanation mechanisms are more user-friendly. AI-driven questions (C5) shows significantly lower ratings ($mean=3.852$, $std=1.799$). It may be because the reflective questions introduce extra cognitive steps before making decisions. Performance visualization (C6) demonstrates a significantly higher score compared to AI-driven questions (C5) ($p=0.017$), suggesting that performance visualization is helpful in guiding user decisions without increasing task complexity.

We find a significant result for \textit{trust and comfort} ($p=0.012$), with textual explanations (C1) scoring higher than human feedback (C4) ($p=0.004$) and AI-driven questions (C5) ($p=0.024$). Human feedback (C4) and AI-driven questions (C5) both result in significantly lower trust scores compared to the more straightforward AI support provided in Condition 1. This indicates that while reflective mechanisms can prompt users to think critically, they may also expose users to uncertainties, reducing trust in the system. Performance visualization (C6) is comparable to other conditions with high trust scores. There are significant differences, with performance visualization (C6) outperforming human feedback (C4) ($p=0.036$) and AI-driven questions (C5) ($p=0.025$), showing that performance visualization is more effective at maintaining user trust.

The \textit{trust dynamics} does not show significant differences across conditions ($p=0.602$). The results indicate that user trust in the AI system remains relatively stable across different explanation mechanisms. However, the mean trust dynamics scores provide some interesting insights into how trust evolved within each condition.
AI CLs (C3) shows the highest trust change ($mean=4.500$, $std=1.607$), suggesting that users recalibrate their trust based on comparing their CL with the AI's CL. Human feedback (C4) has the lowest trust change ($mean=3.444$, $std=1.674$), indicating that inputting CLs lead to a more stable trust level. This may be because users have no information about the AI's certainty, leading them to rely more on their own judgment. Conditions 5 and 6 show moderate trust changes ($mean=4.056$, $std=1.580$ and $mean=3.944$, $std=1.747$), suggesting that users engage with these mechanisms, but do not experience major trust shifts.

\subsection{Impact of Diabetes Management Familiarity on Decision Accuracy}

We divided participants into two groups based on their self-rated familiarity with diabetes management: unfamiliar ($\leq 4$) and familiar ($> 4$). For each condition, we conducted a two-sample t-test to compare decision accuracy between these two groups. The results revealed no statistically significant differences (all $p>0.05$) in average accuracy across conditions. For example, in textual explanations (C1), unfamiliar users ($n=8$) had an average accuracy of $69.4\%$ while familiar users ($n=10$) achieved $72.5\%$ ($p=0.398$). Similarly, in AI CLs (C3) unfamiliar users ($n=11$) achieved $71.8\%$ accuracy compared to $77.9\%$ among familiar users ($p=0.081$). Although this difference trended toward marginal significance, it did not reach the conventional $\alpha = 0.05$ threshold.
These findings suggest that participant familiarity with diabetes management does not systematically impact their overall decision accuracy in each condition. The task was either simple enough for people with general knowledge of healthy eating to complete, or the decision support tools were effective enough to enable users with little prior experience to perform as well as those who were more familiar with the topic—who might have relied more on their own knowledge and experience, rather than the support provided. 

\subsection{Patterns of Trust Evolution Across Conditions}

Our results show some trends in participants' self-reported trust scores across conditions, as shown in Figure~\ref{fig:pre_post_trust}. 
In textual explanation (C1), participants with higher initial trust scores tended to lower their scores in the post-questionnaire, while those with medium initial scores generally increased their trust. A similar pattern was observed in the visual explanation (C2) condition, where where participants with higher initial trust scores were more likely to report reduced trust, as indicated by points below the diagonal line on the pre-questionnaire trust axis. In contrast, AI CLs (C3) demonstrated more consistent scores, with most points remaining on or near the diagonal line, indicating minimal change between pre- and post-questionnaire trust. Both Human Feedback (C4) and AI-Driven Questions (C5) showed declines in trust scores compared to pre-questionnaire levels. The performance visualization (C6) condition revealed a trend where participants with higher initial trust scores tended to report lower trust in the post-questionnaire.
Overall, the trust dynamics across conditions highlight a key trend: participants with higher initial trust were more likely to lower their trust after interacting with the system, especially in conditions where the mechanisms required deeper cognitive engagement (higher EIL). 
This may have occurred because participants with initially higher trust might have had higher expectations of the system. When those expectations were not fully met, either due to system performance or in conditions that required more cognitive effort, their trust may have been undermined. Conversely, participants with medium initial trust levels were more likely to increase their trust, reflecting a potential ``calibration effect'' where interaction with the system helped align their trust levels with their experience of the AI's performance. In no condition, did we observe an increase in trust. It is possible that while AI-generated explanations or CFFs were effective in helping improve human decision-making, they did not necessarily increase trust in the AI. This suggests that users may have treated these explanations as learning tools rather than as mechanisms for trust calibration. Instead of relying more on the AI, users may have developed a stronger mental model, refined their own reasoning process, and become more confident in their independent judgment in the given task. This shift implies that trust in AI may not necessarily grow when users feel empowered to make better decisions with AI's help, though decision-making accuracy may improve. 

\section{Discussion}

\subsection{XAI Methods vs. Cognitive Forcing Functions}
In this section, we analyze the two major areas of decision-support mechanisms in our study: (1) XAI Methods: textual explanations (C1) and visual explanations (C2), and (2) CFFs: AI CLs (C3), human feedback (C4), AI-driven questions (C5), and performance visualization (C6).

\subsubsection{XAI Methods}

Textual Explanations (C1) improve the interpretability of AI's decisions by providing detailed reasoning. It significantly improves user performance and reliability, reflecting the value of detailed reasoning in helping users understand AI outputs. Users generally found the AI's assistance to be helpful. For instance, one user noted, \textit{``The reasoning for its decisions helped me pick which meals were better.''} Several users also reported that the AI expanded their decision-making perspectives, with one stating \textit{``It (the system) helped me come up with ideas and make decisions I wouldn't have made on my own and helped broadened my view.''} Another user similarly remarked, \textit{``It (the system) gave me more perspective of the picture.''} However, this benefit comes with a trade-off. As other studies have noted~\cite{Kaufman2024EffectsOM,fukuchi2024dynamic,bo2024incremental,rong2023towards,bansal2021does}, while textual explanations enhance transparency, they also increase cognitive load. Our findings are consistent with these results; although users appreciate the improved clarity, many find the system more complex to use. A related aspect is how users engage with textual explanations. In our study, participants generally read the explanations provided. However, it is also possible that under time pressure or other circumstances, users might ignore the explanations or fail to engage with them meaningfully \cite{kool2018mental}, necessitating the introduction of CFF methods to enhance engagement with the explanations.

Visual explanations (C2) provide users with an intuitive, graphical breakdown of AI's decision-making process. It does not significantly improve user performance. 
Prior literature has shown mixed results. Some studies have found that visual explanations can provide transparency and improve user satisfaction and trust when well-designed~\cite{8578500,doi:10.1177/1555343411432338,10.1145/2699751}. While other research has highlighted that they can be easily misinterpreted, particularly by lay users, leading to reduced comprehension compared to more detailed textual explanations~\cite{10.1145/3377325.3377480,rago2024exploring}.
Our results align with the latter.
The satisfaction and system complexity remain relatively stable, and the trust does not improve significantly, indicating that visual explanations alone may not be sufficient to enhance deeper engagement or trust, despite the task being visual. This is also reflected in user feedback, with one user noting \textit{"I found that when I chose differently than AI, I sometimes was torn on whether or not I should switch."} Users may need more than visual explanations to meaningfully calibrate trust in AI outputs.

The XAI methods highlight the trade-off between transparency and cognitive load. While the mechanisms can improve user performance and engagement, they should be carefully designed to avoid overwhelming cognitive load.

\subsubsection{Cognitive Forcing Functions}

Recent work emphasizes the critical role of making AI's confidence explicit to enhance human-AI collaboration. Research shows that providing CLs allows users to better assess when and how much to rely on the AI's suggestions~\cite{do2024facilitating,dhuliawala2023diachronic,ma2023should,buccinca2021trust}. 
In our study, we observe a similar pattern: 
AI CLs (C3) significantly improve user performance, showing that making the AI's uncertainty explicit helps users critically engage with its outputs. While trust does not significantly increase, the improved perceived accuracy demonstrates that CLs can act as effective cognitive scaffolds, allowing users to make more informed judgments without feeling cognitively overwhelmed. 
However, the trust does not increase significantly, indicating perhaps the need for more comprehensive explanations to reasonably calibrate user trust. Several users expressed uncertainty regarding the AI's recommendations, with feedback such as, \textit{"I wasn't sure if I should trust the confidence level,"} and \textit{"..., but most of the time I found it hard to believe the AI suggestions."} This feedback reflects a broader issue highlighted in the literature: CLs alone may not suffice for deeper trust calibration. Users may require more explanations to decide whether or not to rely on the AI in complex decision-making scenarios~\cite{li2024overconfident,zhang2020effect}.

Human feedback (C4), especially self-reporting, shows mixed impact on improving trust calibration and decision-making in AI systems. Studies have found that self-reported mechanisms can lead to increased cognitive load~\cite{hoskinghuman,li2024overconfident}. When users are encouraged to reflect on their own confidence~\cite{yeh2001display,rechkemmer2022confidence,pescetelli2021role}, this mechanism can introduce skepticism~\cite{ma2024you}. In our study, human feedback (C4) similarly shows no significant improvements in performance, satisfaction, or trust. One participant stated, \textit{``I had already made my mind up by the third phase, so it didn't play a huge part in my decision-making.''} It highlights how some users may have already discounted AI input when they interact with the feedback mechanism. Another commented, \textit{``I felt like the AI was unreliable.''}, reflecting the observed decline in trust. In essence, while feedback mechanisms encourage active reflection, they must be carefully designed to avoid undermining trust or causing unnecessary complexity. Additionally, when the extra information is provided to the user, it may play a role in whether the user utilizes that information or discounts it, leading to an impact on their perception of the AI's abilities.

Reflective questions have been studied as a tool to engage users more critically with AI outputs~\cite{10.1145/3313831.3376727,shibani2024untangling}, shifting their thinking from fast, heuristic-based decisions (System 1) to more deliberate, analytical processes (System 2)~\cite{kahneman2011thinking}. For example, CFFs show that forcing users to slow down and reflect can reduce over-reliance on AI~\cite{buccinca2021trust,gajos2022people,10.1145/3359204,10.1145/3512930}. However, our results show that while AI-driven questions (C5) encourage engagement, they also reduce trust and reliability. This phenomenon may occur because users become more aware of the AI's potential errors or inconsistencies, as evidenced by participant feedback. One user remarked, \textit{"It felt a bit unreliable when it kept saying that there was pizza on the plate despite there very clearly being no pizza on the plate."} Similarly, another noted, \textit{"When I saw the AI incorrectly thought one plate had pizza, it made me want to pick the other one instead."}
Despite the decrease in trust, we observe a significant reduction in system complexity, indicating that reflective questions may streamline the decision-making process by focusing user attention on key aspects. However, it comes at the cost of trust, particularly when AI errors are highlighted during the reflective process. Designing such questions to avoid undermining trust remains a challenge, but when done well, reflective engagement can support a more thoughtful interaction between humans and AI systems.

Performance visualization (C6) has been studied as a powerful mechanism for enhancing human-AI interactions. Research shows that visual aids can help users form mental models of AI behavior, influencing their trust and engagement levels. Such mechanisms help users justify their decisions in AI-assisted tasks by making the AI's behavior more transparent~\cite{hoque2024harder,Raees2024FromET,gomez2023designing,cabrera2023improving,ferreira2021human,Ramos2020InteractiveMT}. Performance visualization (C6) in our study demonstrates borderline significance in improving user performance, indicating that this mechanism may positively influence users to critically engage with the task. The improvements suggest that users reflected more deeply on the comparison between their own decisions and the AI's, as evidenced by remarks like, \textit{"I used it to reflect"} and \textit{"The bar chart helped give me a visual 'faith' in the AI."} 
The mechanism serves as a cognitive aid, helping users refine their judgments through the visual representation of performance. However, the lack of stronger performance improvements shows it can be enhanced with more actionable insights to further improve the decision-making outputs. This aligns with previous works, which emphasize visual aids must be carefully designed to prevent overwhelming users or failing to deliver impactful insights~\cite{10.1145/3290605.3300809,Guo2020SurveyOV,Bylinskii2017LearningVI}.

\subsection{Under-reliance and Over-reliance on AI}
We explore the phenomena of under-reliance and over-reliance when humans interact with AI systems. We analyze how reliance evolves between two stages of interaction: Phase 2 (AI support without the designed AI support mechanisms) and Phase 3 (AI support with the additional mechanism). To quantify the change in reliance, we measure the extent to which users follow AI suggestions when the AI is correct, compared to when the AI is incorrect. The results provide insights into how decision support mechanisms influence trust or reliance calibration, highlighting both under-reliance (insufficient reliance on correct AI suggestions) and over-reliance (excessive reliance on incorrect AI suggestions).

Textual explanations (C1) have a minor, non-significant increase in reliance calibration from Phase 2 to Phase 3. The relatively stable reliance dynamics suggest that while users find the detailed reasoning helpful, the explanations do not immediately shift their reliance on the AI. The textual explanations likely contribute to maintaining a steady level of reliance, avoiding dramatic swings toward under-reliance or over-reliance~\cite{10.1145/3397481.3450650,10.1145/3377325.3377480}.

In visual explanations (C2), reliance calibration slightly increases from Phase 2 to Phase 3. It suggests that visual explanations can help users better align with AI, but the change is not statistically significant. Such a modest increase implies that while users find visual explanations somewhat useful, they are not sufficient to help users identify when the AI is wrong, as users may rely on the visual aids without critically evaluating whether the AI's suggestions are accurate~\cite{xuan2023can,chu2020visual}.

In AI CLs (C3), we observe a significant increase in reliance from Phase 2 to Phase 3. The transparency of the AI's confidence allows users to build reasonable reliance, and have a better alignment with the AI when it is correct and reduce reliance when it is incorrect. 
The significant improvement highlights the importance of providing quantifiable measures of AI certainty to build more accurate reliance, minimizing both under-reliance and over-reliance~\cite{cao2024designing,ma2023should,rauker2023toward}.

In human feedback (C4), reliance calibration remains relatively stable. The lack of significant change suggests that while reflecting on their own confidence may engage users more critically, it does not substantially improve their ability to align with AI outputs. Reliance calibration in this condition appears to be less effective, as users may struggle to map their confidence with the AI's accuracy~\cite{labarta2024study}, leading to minimal impact.

AI-driven questions (C5) led to a significant improvement in reliance calibration from Phase 2 to Phase 3, showing users are more likely to follow the AI's suggestions when they are correct and less likely to follow them when they are incorrect.
The reflective questions prompt users to think more critically about the AI's suggestions, engaging them to question the AI's reasoning~\cite{glinka2023critical}. However, it is essential to balance the cognitive load, as questions may introduce complexity that could lead to overthinking or under-reliance~\cite{10.1145/3613904.3642902}.

In performance visualization (C6), we find a slight decrease in reliance calibration from Phase 2 to Phase 3, though the change is not statistically significant. 
The results suggest that while users appreciate the comparison information, it does not necessarily lead to improved reliance calibration. Users still exhibit a tendency to follow the AI even when it is wrong. Visualizing past performance may introduce additional complexity without direct model interpretability, which likely leads users to rely more on the AI, even when it is incorrect.

\subsubsection{Mitigating Over-reliance Through Transparency and Reflection}

Across these conditions, we observe mechanisms aimed at transparency and interpretability (textual explanations (C1) and AI CLs (C3)) and reflective engagement (AI-driven questions (C5)) are the most effective at mitigating over-reliance when the AI is incorrect. These mechanisms help users engage critically with the AI's outputs, reducing the likelihood of blindly following incorrect suggestions. On the other hand, mechanisms such as visual explanations and performance visualizations may not provide sufficient guidance, leading to a higher risk of over-reliance when the AI has errors. The results highlight the importance of reliance calibration mechanisms that go beyond surface-level transparency, offering users clear, actionable information or opportunities for active reflection.

\subsection{Balancing Cognitive Load and Reliance Calibration in AI-Assisted Decision-Making}

Our results highlight the critical balance between cognitive load management and reliance calibration in human-AI collaboration. As AI systems become more integrated into high-stakes domains, understanding the dynamics of under-reliance and over-reliance is essential for designing effective AI support mechanisms. We discuss key findings of previous research and the implications for future AI system design in this subsection.

\subsubsection{Cognitive Load Management}
Our findings support the hypothesis that \textit{balanced EIL is crucial for optimizing human-AI collaboration (H3)}, especially in high-stakes environments like diabetes management. 
Conditions involving textual explanations (C1) and AI CLs (C3) show significant improvements in user accuracy and decision-making, as these mechanisms provide transparency and help users engage critically with AI outputs without overwhelming them. 
Previous research has similarly demonstrated that well-calibrated transparency and interpretability can build more effective reliance calibration, allowing users to rely on AI outputs more appropriately without falling into over-reliance or under-reliance traps~\cite{cao2024designing,poursabzi2021manipulating, bansal2021does,liu2021understanding,lai2019human}.

Mechanisms such as performance visualizations (C6) provide valuable historical insights but lead to borderline significant improvements in user accuracy. While users benefit from seeing comparison data, the additional cognitive load required to process this information seems to reduce the overall effectiveness of reliance calibration. These findings align with cognitive psychology studies, where increased reflection and mental effort, when not accompanied by clear decision-making guidance, can lead to cognitive overload and hinder performance~\cite{darejeh2024critical,ashktorab2024emerging, lebiere2021adaptive}.

\subsubsection{The Importance of Reliance Calibration Mechanisms}

Our analysis of reliance dynamics shows the complexity of reliance calibration, especially, when AI provides an incorrect output.  
In AI CLs (C3), we observe significant improvements in user ability to calibrate reliance correctly, aligning their responses more closely with correct AI outputs and disengaging from incorrect suggestions. It shows the importance of quantifiable metrics, which provide direct insight into AI certainty~\cite{zhang2024evaluating,logg2019algorithm}. As noted in prior work, AI systems that explicitly communicate uncertainty can help users calibrate their reliance on AI outputs, fostering more accurate decision-making~\cite{marusichusing}.

However, in visual explanations (C2) and performance visualizations (C6), users find it difficult to disengage from incorrect AI outputs, leading to over-reliance. The results suggest that surface-level transparency without deeper context may not be enough to guide users in critical decision-making scenarios. Users tend to over-rely on automation~\cite{zhai2024effects,vasconcelos2023explanations,logg2019algorithm,goddard2012automation}, especially when the decision-making process lacks sufficient transparency or when the cognitive effort to process the decision information is high.

Moreover, human feedback (C4) is designed to encourage user reflection and active engagement with AI outputs. It does not lead to better reliance calibration or decision accuracy. The results are supported by previous research, which suggests that reflective mechanisms may not always lead to positive outcomes, and users may struggle with comparative reflection in high-cognitive tasks~\cite{buccinca2021trust,Bansal_Nushi_Kamar_Lasecki_Weld_Horvitz_2019,schworm2007learning}.

Interestingly, the results for AI-driven questions (C5)
demonstrate reduced reliance calibration but improved decision-making accuracy. The cognitive load imposed by reflective questions may lead users to engage more critically with AI outputs, but it also introduces skepticism. Users find it more challenging to fully trust the AI's reasoning~\cite{tankelevitch2024metacognitive, buccinca2021trust}. It shows a key tension in AI design: while reflective engagement can promote deeper critical thinking, it must be carefully calibrated to prevent over-scrutiny~\cite{kazemitabaar2024exploring,naik2024generating}.

\section{Limitations and Future Work}
While our study provides valuable insights into the effectiveness of different interface design mechanisms in influencing user trust, engagement, and decision-making, several limitations should be considered to provide a comprehensive understanding of the results and their generalizability. 
First, the study focused on diabetes management, specifically meal choices, which is a highly specific and high-stakes domain. The findings may not fully generalize to other decision-making domains like finance, law, or education. 
Different domains come with unique cognitive demands and stakes, which can influence how users engage with AI. Future studies should consider evaluations of the proposed mechanisms in varied domains to understand the broader applicability of our findings. 
Second, we examined reliance calibration over relatively short-term interactions within a controlled experimental setup. Although it is a typical approach~\cite{buccinca2021trust,ma2023should}, it may not fully capture long-term dynamics. The real-world human-AI collaboration often involves evolving interactions where reliance dynamics can shift over time~\cite{xu2023comparing}. Users may initially over-rely or under-rely on the AI, but these behaviors may change as they interact with the system over an extended period and depend on how much value the system provides them. Future studies could consider longitudinal studies to understand how reliance evolves and whether AI support mechanisms need to adapt dynamically as users become more familiar with the system.

\section{Conclusion}
Our study investigates the impact of six decision-support mechanisms on user trust, engagement, and decision-making in high-stakes scenarios like diabetes management. Each mechanism is designed to emphasize specific aspects of the decision-making process, targeting either System 1 (fast, intuitive, heuristics driven, prone to cognitive biases) or System 2 (deliberative, analytical, effortful, focused on logical reasoning) cognitive processes. Our findings reveal the following dynamics: explanation mechanisms that facilitate transparent, interpretable reasoning, such as textual explanations and AI CLs, improve decision-making accuracy by helping users better align their choices with AI suggestions. However, CFF mechanisms like human feedback and AI-driven questions, while encouraging deeper engagement, sometimes impose additional cognitive demands, thereby reducing performance. Importantly, none of the mechanisms simultaneously support both System 1 and System 2 thinking, highlighting the trade-offs between intuitive usability and reflective deliberation in AI interface design. Visual explanations and performance visualizations offer straightforward support with low-cognitive-load but are less effective in fostering critical evaluation of AI outputs and therefore have limited impact on trust and reliance. In contrast, mechanisms encouraging reflection promote deeper analysis but risk overwhelming users when the cognitive load is too high.

Our results highlight the need for adaptive decision-support mechanisms, including task complexity, user expertise, and decision frequency. Our findings emphasize that a balanced design approach, where decision-support mechanisms are matched with user cognitive capacity and task demands, is essential for optimizing human-AI collaboration. Specifically, mechanisms with manageable EIL and interpretable outputs demonstrated the best balance between fostering engagement and maintaining performance.

Our future research will explore adaptive interfaces capable of dynamically adjusting explanation complexity based on user behavior and task difficulty. Longitudinal studies are also needed to examine how reliance on AI evolves over time and across domains. By advancing our understanding of these dynamics, we can design AI systems that not only improve decision-making accuracy but also support safe and sustainable human-AI collaboration in high-stakes environments.

\section{Methods} \label{methods}
\begin{figure}[htbp]
    \centering
    \includegraphics[width=0.9\linewidth]{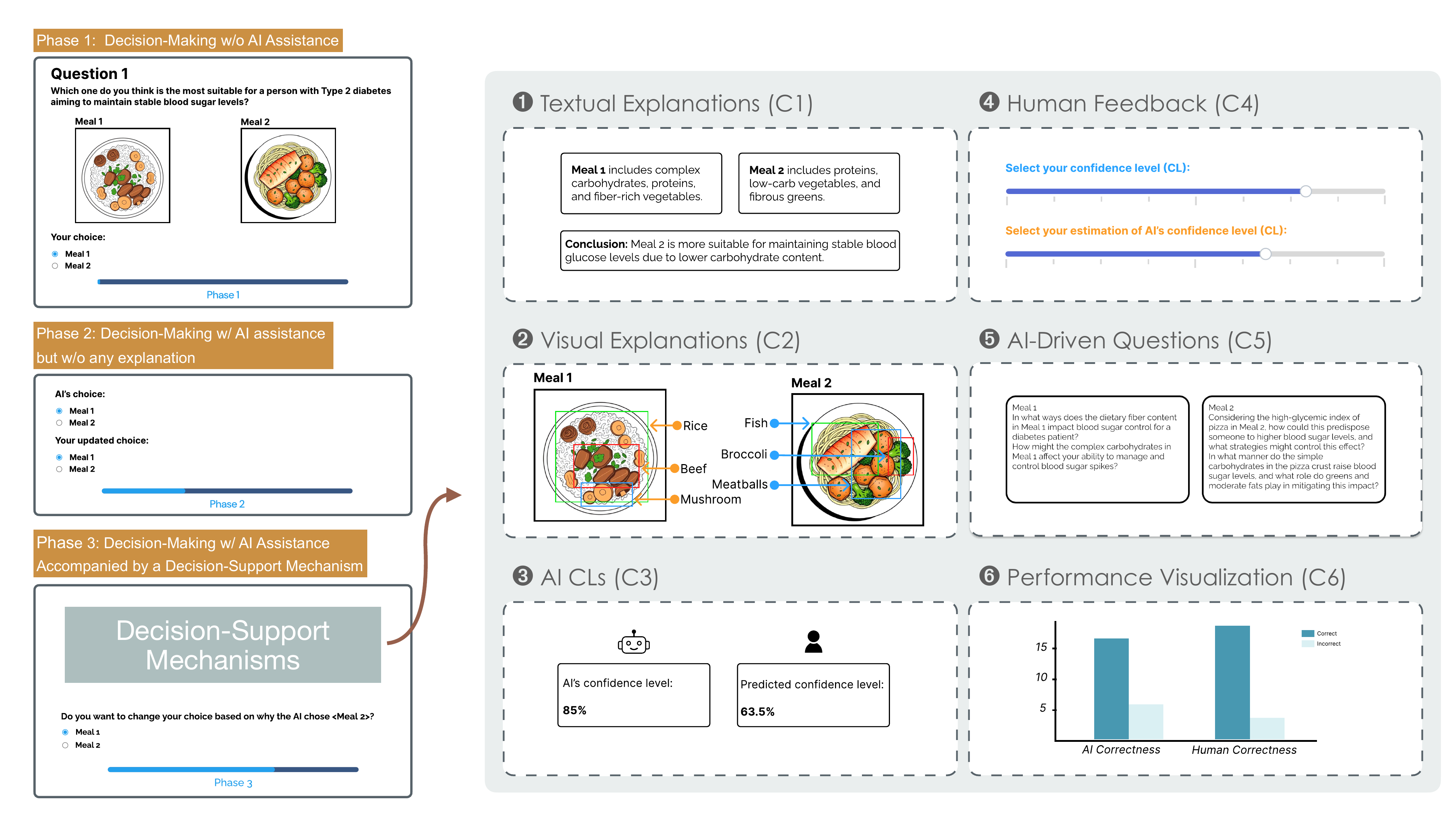}
    \caption{Interface design across different phases: Phase 1, where the user makes an independent decision without AI assistance; Phase 2, where the AI presents its suggestion but without any explanation; and Phase 3, where condition-specific AI decision-making support mechanisms are applied, followed by another chance for users to update their decision. The right side of the figure illustrates the various explanation methods employed. (1) Textual explanation for why the AI selected a preferred meal, (2) Visual segmentation and labeling of meal items, (3) AI's confidence level (CL) and an estimation the user's confidence level (CL), (4) Human feedback on the AI and their own confidence levels (CLs), (5) AI prompted questions for the user, and (6) Performance visualization of the correctness of past questions, for both the AI and the user.}
    \label{fig:interface}
\end{figure}

Our experiments were conducted using a custom web-based interface (Figure~\ref{fig:interface}), designed to simulate the interaction between users and a generative AI system tasked with assisting in meal planning for diabetes management. Participants were presented with an experimental condition that represented an AI explanation mechanism. Each condition consisted of three phases.

\subsection{Participants}
We recruited 108 participants (N=108) for this study through Prolific\footnote{\url{https://www.prolific.com/}}, each assigned randomly to one of six experimental conditions, with 18 participants per condition. 
Before data collection, we performed a power analysis~\cite{cohen2013statistical} to ensure that our design would be adequately powered for within-group and between-group analyses. Specifically, an effect size of $r=0.60$ at $80\%$ power and $\alpha=0.05$ requires approximately 18 participants. Thus, having 18 participants per condition was sufficient to detect significant effects. 
Participants were drawn from a diverse pool, ensuring variation in age, gender, educational background, familiarity with AI systems, and knowledge of diabetes management. We set primary language as English, and participants had a Prolific approval rating of 99-100\%, demonstrating a high level of engagement and reliable participation history. Participants were compensated at an hourly rate of \$10.15, with an additional \$1 bonus for those whose decision accuracy exceeded 75\%. 
Additionally, participants were screened to ensure that they had no medical expertise in diabetes management to simulate the experience of general users engaging with AI systems in healthcare contexts. The overview of participant demographic information and familiarity with diabetes is shown in Table~\ref{tab:demographics}.
Informed consent was obtained from all participants prior to the experiment, following institutional ethical guidelines.

\begin{table}[htbp]
\centering
\begin{tabular}{lll}
\hline
\textbf{Measure}             & \textbf{Statistics}         & \textbf{Value}      \\ \hline
\textbf{Familiar with Diabetes} & Min                        & 1                  \\ 
                                & Max                        & 7                  \\   
                                & Mean                       & 4.09               \\   
                                & Median                     & 4.0                \\   
                                & Std          & 1.86               \\ \hline
\textbf{Age}                  & Min                        & 16                 \\   
                                & Max                        & 67                 \\   
                                & Mean                       & 36.99              \\   
                                & Median                     & 35.5               \\   
                                & Std          & 12.01              \\ \hline
\textbf{Education}            & High school degree          & 53.78\%                 \\   
                                & College degree             & 38.89\%                 \\   
                                & Master's degree            & 4.63\%                  \\   
                                & PhD degree                 & 3.70\%                  \\ \hline
\textbf{Gender}               & Woman                      & 51.85\%                 \\   
                                & Man                        & 45.37\%                 \\   
                                & Nonbinary                  & 1.85\%                  \\   
                                & Other                      & 0.93\%                  \\ \hline
\end{tabular}
\vspace{5pt}
\caption{Summary of participant demographic information and familiarity with diabetes.}
\label{tab:demographics}
\end{table}

\subsection{Task and Procedure}

Participants were asked to perform a simulated meal-planning task. The task was designed to reflect real-world decisions in diabetes management. The task was divided into three phases across six conditions:
\begin{enumerate}
    \item \textbf{Phase 1: Decision-Making w/o AI Assistance \space\space}
        Participants were presented with two meal options and asked to choose which meal would be most suitable for maintaining stable blood sugar levels for a Type 2 diabetic person. At this stage, participants made decisions without AI assistance. This constituted a baseline measure of their own decision-making accuracy.
    \item \textbf{Phase 2: AI Assistance (AI-Only)}
        Participants were presented with the same meal options as in Phase 1, but they received AI assistance for meal recommendation at this stage. However, no explanation was provided for the AI's decision. This phase was designed to measure the impact of AI assistance on user decision-making accuracy.
    \item \textbf{Phase 3: Condition-Specific AI Explanation Mechanism}
        Participants were shown one of six different AI explanation mechanisms. Following this new information, participants could either revise or affirm their previous decision. This phase was designed to measure the impact of proposed AI explanation mechanisms on user decision-making.

\end{enumerate}

\subsection{Experimental Conditions}
The conditions were designed to evaluate the effectiveness of AI explanation mechanisms in influencing user trust, engagement, and decision-making. Below, we describe each condition, including the rationale for its design and the expected impact on user trust and engagement.

\begin{enumerate}
    \item \textbf{Textual Explanations (C1):}
        \begin{itemize}
            \item Condition Content: Textual explanation provides the rationale behind the AI's reasoning (e.g., ``Meal 2 contains fewer carbohydrates, which is critical for managing blood sugar levels''). 
            \item Objective: To test whether a logical, written explanation increases transparency and promotes critical thinking, thereby influencing trust calibration.
        \end{itemize}
    \item \textbf{Visual Explanations (C2):}
        \begin{itemize}
            \item Condition Content: Visual explanation highlights the key nutritional components of the meal (e.g., carbohydrates, proteins), and is provided with a concise nutritional content of the meal. This graphical representation is designed to intuitively communicate the rationale behind the AI's suggestion.
            \item Objective: To test whether visual explanations help reduce cognitive load and improve user trust by providing easily interpretable information.
        \end{itemize}
    \item \textbf{AI CLs (C3)):}
        \begin{itemize}
            \item Condition Content: Confidence levels are provided for AI and users. AI's CL is calculated by the model's uncertainty in the prediction. User's CL is calculated based on the user's history of current task performance. The detailed calculation is provided in the Section~\ref{cal_prompt}: \textit{Prompt for Calculating User’s CL}.
            \item  Objective: To test whether providing CLs for both AI and users helps users better calibrate their trust, offering a quantifiable metric to assess the AI's certainty and the user's confidence on the same scale.
        \end{itemize}
    \item \textbf{Human Feedback (C4):}
        \begin{itemize}
            \item Condition Content: Participants input their CLs for both AI and themselves. 
            \item Objective: To test whether human feedback can help users engage with the AI system, by reflecting on their own CL and AI's CL, and calibrating their trust accordingly.
        \end{itemize}
    \item \textbf{AI-Driven Questions (C5):}
        \begin{itemize}
            \item Condition Content: Three critical questions are asked to the user, which are designed to help users reflect on their decision-making process and the AI's recommendation.
            \item Objective: To test whether prompting users to reflect on their decision-making process and the AI's recommendation can help users engage with the AI system deeply and calibrate their trust.
        \end{itemize}
    \item \textbf{Performance Visualization (C6):}
        \begin{itemize}
            \item Condition Content: Participants were shown a performance visualization graph, comparing the AI's performance over past decisions with their own.
            \item Objective: To test whether visualizing the AI's performance over past decisions can help users calibrate their trust and engage with the AI system.
        \end{itemize}
    \end{enumerate}

\subsubsection{Prompt Design}
The following prompts are designed to align with the objectives of each condition, ensuring that the AI output supports the rationale behind the design of the condition.

\definecolor{softGray}{RGB}{240, 240, 240}  
\definecolor{deepBlue}{RGB}{0, 76, 153}     

\begin{tcolorbox}[colback=softGray, colframe=deepBlue, title=System Prompt, 
fonttitle=\fontsize{8pt}{1pt}\selectfont, fontupper=\fontsize{8pt}{1pt}\selectfont,]

You are a nutritionist specializing in diabetes management. You are asked to provide a concise answer to the most suitable meal for managing blood sugar levels for a diabetes patient.

\end{tcolorbox}

\begin{tcolorbox}[colback=softGray, colframe=deepBlue, title=Textual Explanation (C1), 
fonttitle=\fontsize{8pt}{1pt}\selectfont, fontupper=\fontsize{8pt}{1pt}\selectfont,]

Given two images of meals, your task is to analyze and compare their potential impact on blood sugar levels for a diabetes patient. Assess the visible foods for their carbohydrate types (simple vs. complex), presence of dietary fiber, and overall balance of macronutrients (carbohydrates, proteins, fats). Identify any high-glycemic ingredients and estimate the balance of the meals in terms of diabetes-friendly nutrition.

Only provide a very concise explanation of reasoning on which meal might be more suitable for maintaining stable blood glucose levels. Do not provide any detailed ingredients of the meal, only mention macronutrients. You need to mention ``Meal 1'' and ``Meal 2,'' and have a concise conclusion in your answer. 

\end{tcolorbox}

\begin{tcolorbox}[colback=softGray, colframe=deepBlue, title=Visual Explanation (C2), 
fonttitle=\fontsize{8pt}{1pt}\selectfont, fontupper=\fontsize{8pt}{1pt}\selectfont,]

Given two images of meals, your task is to analyze and compare their potential impact on blood sugar levels for a diabetes patient. Assess the visible foods for their carbohydrate types (simple vs. complex), presence of dietary fiber, and overall balance of macronutrients (carbohydrates, proteins, fats). Identify any high-glycemic ingredients and estimate the balance of the meals in terms of diabetes-friendly nutrition.

Based on the meal images and its segmentation, think about calorie and nutritional content, please provide the number of the meal that is a better choice for managing blood sugar levels for a diabetes patient, followed by a list of the foods identified by the numbered boxes in each image, with no additional details or explanations. Present the information in the following format:

\texttt{Better option: [Image number]} 

\texttt{First Meal: list of food by number} 

\texttt{Second Meal: list of food by number}
\end{tcolorbox}

\begin{tcolorbox}[colback=softGray, colframe=deepBlue, title=AI CL (C3) and Performance Visualization (C6), 
fonttitle=\fontsize{8pt}{1pt}\selectfont, fontupper=\fontsize{8pt}{1pt}\selectfont,]

Given two images of meals, your task is to analyze and compare their potential impact on blood sugar levels for a diabetes patient. Assess the visible foods for their carbohydrate types (simple vs. complex), presence of dietary fiber, and overall balance of macronutrients (carbohydrates, proteins, fats). Identify any high-glycemic ingredients and estimate the balance of the meals in terms of diabetes-friendly nutrition.

Choose the meal that is more suitable for a diabetes patient based on the nutritional content and potential impact on blood sugar levels. Provide only the image number (either ``Meal 1'' or ``Meal 2'') and your confidence level in the assessment as a percentage from 0 to 100. Do not provide any additional explanation.

\end{tcolorbox}

\begin{tcolorbox}[colback=softGray, colframe=deepBlue, title=AI-driven Questions (C5), 
fonttitle=\fontsize{8pt}{1pt}\selectfont, fontupper=\fontsize{8pt}{1pt}\selectfont,]

Given the description of two meals, generate questions. These questions should explore how these components impact, control, or raise blood sugar levels. Provide two questions per meal, using different starting phrases and using the verbs ``control'', ``impact'', and ``raise'' to focus on blood sugar management. The description of the meals is as follows:
[MEAL\_DESCRIPTION].

\end{tcolorbox}

\begin{tcolorbox}[colback=softGray, colframe=deepBlue, title=Prompt for Calculating User's CL, 
fonttitle=\fontsize{8pt}{1pt}\selectfont, fontupper=\fontsize{8pt}{1pt}\selectfont, label=cal_prompt]

Based on historical data of \texttt{\{n\_records\}} user decisions on meal choices for controlling blood sugar levels for diabetes patients, predict the user’s confidence in choosing the correct meal for their next query. Use the nutritional profiles and the user’s past choices to model behavior. Below is the historical data and the query for prediction:

\textbf{User Historical Data Points:}

\texttt{\{For each record\}} (Record \texttt{{i+1}}):

    - Meal 1 Meta Info:
    
      Calories: \texttt{\{meal1\_calories\}}. Total Mass: \texttt{\{meal1\_mass\}}. Total Fat: \texttt{\{meal1\_fat\}}. Total Carbohydrates: \texttt{\{meal1\_carbohydrates\}}. Total Protein: \texttt{\{meal1\_protein\}}  
      
    - Meal 2 Meta Info:
    
      Calories: \texttt{\{meal2\_calories\}}. Total Mass: \texttt{\{meal2\_mass\}}. Total Fat: \texttt{\{meal2\_fat\}}. Total Carbohydrates: \texttt{\{meal2\_carbohydrates\}}. Total Protein: \texttt{\{meal2\_protein\}}. Ground Truth: \texttt{\{ground\_truth\}}. User's Choice: \texttt{\{user\_choice\}}  

\textbf{New Query for Prediction:}

- \textbf{Meal 1 Meta Info:}

    Calories: \texttt{\{new\_meal1\_calories\}}. Total Mass: \texttt{\{new\_meal1\_mass\}}. Total Fat: \texttt{\{new\_meal1\_fat\}}. Total Carbohydrates: \texttt{\{new\_meal1\_carbohydrates\}}. Total Protein: \texttt{\{new\_meal1\_protein\}} 
  
- \textbf{Meal 2 Meta Info:}

  Calories: \texttt{\{new\_meal2\_calories\}}. Total Mass: \texttt{\{new\_meal2\_mass\}}. Total Fat: \texttt{\{new\_meal2\_fat\}}. Total Carbohydrates: \texttt{\{new\_meal2\_carbohydrates\}}. Total Protein: \texttt{\{new\_meal2\_protein\}}. The ground truth is: \texttt{\{new\_ground\_truth\}}.

\textbf{Instruction:}  
Return the predicted likelihood as a percentage (0-100) of the user choosing the correct meal based on their past behavior and the nutritional profile needed for controlling blood sugar level goals for diabetes patients. Do not provide any additional explanation.

\end{tcolorbox}

\subsection{Pre- and Post-study Questionnaires}
Before and after the study, participants were asked to respond to a series of questionnaires. The pre-study questionnaire is used to gather background information on participant AI usage and their baseline levels of trust in AI systems. The post-study questionnaire is used to assess participant experience with proposed AI explanation mechanisms. The detailed questions are provided in Section ~\ref{appdeix:measures}.



\subsection{Statistical Analysis}

We employ statistical methods, adapted to different comparisons, to evaluate the impact of AI explanation mechanisms. These analyses are designed to test our hypotheses regarding user behavior and performance under different conditions.
\begin{enumerate}
    \item Reliance on AI Suggestions (H1): To investigate user reliance on AI suggestions, we apply the McNemar test~\cite{mcnemar1947note}. This test is appropriate for paired nominal data, enabling us to evaluate whether the proportion of users aligning with AI suggestions differed significantly when the AI's suggestions are correct versus incorrect. The McNemar test provides insights into the degree of over- (or under-) reliance by analyzing changes in user behavior based on the accuracy of the AI's outputs.

    \item Within-Condition Comparisons (H2, H3): To examine differences between the three phases within each condition, we first assessed data normality with the Shapiro-Wilk test~\cite{10.1093/biomet/52.3-4.591}. Textual explanations (C1) ($p=0.004$), visual explanations (C2) ($p=0.001$), AI CLs (C3) ($p<0.0001$) and performance visualizations (C6) ($p=0.045$) did not meet normality assumptions, while human feedback (C4) ($p=0.051$) and AI-driven questions (C5) ($p=0.061$) were borderline above $0.05$. Given these mixed or borderline results, we use the Wilcoxon signed-rank test~\cite{wilcoxon1992individual}. This non-parametric test is suitable for paired data and does not assume a normal distribution, making it appropriate for analyzing the non-normally distributed data. The Wilcoxon signed-rank test allows us to assess whether there are significant improvements in user performance (e.g., decision accuracy), trust, or engagement across the three phases and between pre- and post-questionnaires within each condition.

    \item Between-Condition Comparisons (H3): 
    To compare the effects of different AI explanation mechanisms across conditions, we employ the Kruskal-Wallis test~\cite{kruskal1952use}, a non-parametric alternative to one-way ANOVA. This method is appropriate for comparing independent groups with non-normally distributed data. By applying the Kruskal-Wallis test, we evaluate whether the type of AI explanation mechanism significantly influences user performance, trust, and engagement.

\end{enumerate}

\subsection{Code Availability}

All statistical analyses are conducted using Python 3.10.0 and the SciPy library. The significant threshold is set at $p < 0.05$ for all statistical tests. All data and code will be made available upon paper publication.

\subsection{Measures}\label{appdeix:measures}

\subsubsection{Data Collection}

Participant responses were recorded at each phase of the task, with data collection focusing on multiple dependent variables designed to capture user performance, trust, and engagement: 

\begin{enumerate}
    \item User Performance (Accuracy): The correctness of participant choices relative to optimal choices for diabetes management, as determined by the ground truth (GT). The GT was computed based on guidelines from the American Diabetes Association, detailed in Section~\ref{sec:gt}.
    \item Pre-Post Questionnaire Comparisons: 
        \begin{itemize}
            \item Satisfaction: User reported satisfaction with the AI-assisted decision-making process.
            \item System Complexity: Perceived complexity of interacting with the AI system.
            \item Reliability: User perception of the AI's reliability.
            \item Trust: Previous experience and post-task trust in AI systems.
            \item System Accuracy: Users' perceived accuracy of the AI in providing correct suggestions.
        \end{itemize}
    \item Post-Task Questionnaire Measures:
        \begin{itemize}
            \item Confidence and Understanding: User confidence in their decisions and understanding of the AI's suggestions.
            \item Consistency and Cognitive Load: Perceived consistency in the AI's outputs and the cognitive effort required to process its explanations.
            \item Ease and Usefulness: The ease of using the AI system and the perceived utility of its suggestions in decision-making.
            \item Trust and Comfort: Levels of trust in the AI's outputs and overall comfort with relying on the system.
            \item Trust Dynamics: Changes in user trust in the AI from the beginning to the end of the task.
        \end{itemize}
\end{enumerate}


Each participant completed a pre-questionnaire to assess their familiarity with AI systems and baseline trust, followed by a post-questionnaire evaluating their satisfaction, trust dynamics, and perceived interaction quality after engaging with the AI-assisted decision-making process.

\subsubsection{Data and Statistical Analysis}

\begin{table}[ht]
\centering
\begin{tabular}{l|c|c|c}
\hline
\textbf{Measure} & \textbf{Condition} & \textbf{Mean (SD)} & \textbf{Significant Differences} \\
\hline
\multirow{6}{*}{\makecell[l]{Confidence and\\Understanding\\($H = 16.11$,\\$p = 0.007^*$)}} 
& Textual Explanations (C1) & 5.444 (0.970) & \multirow{6}{*}{\makecell[l]{1 > 4 ($p = 0.002^*$)\\1 > 5 ($p = 0.008^*$)\\1 > 6 ($p = 0.003^*$)}} \\
& Visual Explanations (C2) & 6.000 (0.553) & \\
& AI CLs (C3) & 5.528 (0.889) & \\
& Human Feedback (C4) & 4.917 (1.182) & \\
& AI-Driven Questions (C5) & 5.250 (0.854) & \\
& Performance Visualization (C6) & 4.583 (1.465) & \\
\hline
\multirow{6}{*}{\makecell[l]{Consistency and\\Cognitive Load\\($H = 6.48$,\\$p = 0.262$)}} 
& Textual Explanations (C1) & 4.639 (0.760) & \multirow{6}{*}{None} \\
& Visual Explanations (C2) & 4.667 (0.913) & \\
& AI CLs (C3) & 4.694 (1.056) & \\
& Human Feedback (C4) & 4.111 (1.264) & \\
& AI-Driven Questions (C5) & 4.306 (0.974) & \\
& Performance Visualization (C6) & 3.944 (1.212) & \\
\hline
\multirow{6}{*}{\makecell[l]{Ease and\\Usefulness\\($H = 12.44$,\\$p = 0.029^*$)}} 
& Textual Explanations (C1) & 5.204 (1.428) & \multirow{6}{*}{\makecell[l]{1 > 4 ($p = 0.025^*$)\\1 > 6 ($p = 0.009^*$)\\2 > 6 ($p = 0.033^*$)\\5 > 6 ($p = 0.017^*$)}} \\
& Visual Explanations (C2) & 5.444 (0.882) & \\
& AI CLs (C3) & 4.944 (1.268) & \\
& Human Feedback (C4) & 4.111 (1.802) & \\
& AI-Driven Questions (C5) & 5.315 (1.057) & \\
& Performance Visualization (C6) & 3.852 (1.799) & \\
\hline
\multirow{6}{*}{\makecell[l]{Trust and\\Comfort\\($H = 14.62$,\\$p = 0.012^*$)}} 
& Textual Explanations (C1) & 5.111 (1.242) & \multirow{6}{*}{\makecell[l]{1 > 4 ($p = 0.004^*$)\\1 > 6 ($p = 0.024^*$)\\3 > 4 ($p = 0.014^*$)\\3 > 6 ($p = 0.029^*$)\\5 > 4 ($p = 0.036^*$)\\5 > 6 ($p = 0.025^*$)}} \\
& Visual Explanations (C2) & 5.426 (0.800) & \\
& AI CLs (C3) & 5.241 (0.859) & \\
& Human Feedback (C4) & 4.185 (1.450) & \\
& AI-Driven Questions (C5) & 5.241 (1.076) & \\
& Performance Visualization (C6) & 4.037 (1.717) & \\
\hline
\multirow{6}{*}{\makecell[l]{Trust\\Dynamics\\($H = 3.64$,\\$p = 0.602$)}} 
& Textual Explanations (C1) & 4.167 (1.740) & \multirow{6}{*}{None} \\
& Visual Explanations (C2) & 4.167 (1.572) & \\
& AI CLs (C3) & 4.500 (1.607) & \\
& Human Feedback (C4) & 3.444 (1.674) & \\
& AI-Driven Questions (C5) & 3.944 (1.747) & \\
& Performance Visualization (C6) & 4.056 (1.580) & \\
\hline
\end{tabular}
\vspace{5pt}
\caption{Statistical comparison of measures across conditions. $H$ represents the Kruskal-Wallis test statistic. Asterisks ($*$) indicate significant results ($p < 0.05$). For significant differences, the ``$>$'' symbol indicates that the first condition had a significantly higher score than the second condition.}
\label{tab:condition-comparisons}
\end{table}

\begin{table}[ht]
\centering
\begin{tabular}{l|cc|cc|c}
\hline
\multirow{2}{*}{\textbf{Condition}} & \multicolumn{2}{c|}{\textbf{P2 Change}} & \multicolumn{2}{c|}{\textbf{P3 Change}} & \multirow{2}{*}{\textbf{Significance}} \\
\cline{2-5}
& Mean & SD & Mean & SD & \\
\hline
Textual Explanations (C1) & 0.139 & 0.101 & 0.067 & 0.090 & $p = 0.049^*$ \\
Visual Explanations (C2) & 0.100 & 0.097 & 0.108 & 0.089 & $p = 0.776$ \\
AI CLs (C3) & 0.183 & 0.129 & 0.033 & 0.044 & $p = 0.001^*$ \\
Human Feedback (C4) & 0.072 & 0.079 & 0.006 & 0.016 & $p = 0.006^*$ \\
AI-Driven Questions (C5) & 0.092 & 0.069 & 0.044 & 0.052 & $p = 0.058$ \\
Performance Visualization (C6) & 0.456 & 0.108 & 0.394 & 0.119 & $p = 0.009^*$ \\
\hline
\end{tabular}
\vspace{5pt}
\caption{Comparison of engagement changes between P1-P2 (P2 Change) and P1-P3 (P3 Change) within each condition. Asterisks ($*$) indicate significant differences ($p < 0.05$) between P2 and P3 changes based on Wilcoxon signed-rank tests.}
\label{tab:engagement-changes}
\end{table}

\begin{figure}[ht]
\centering
\includegraphics[width=0.7\textwidth]{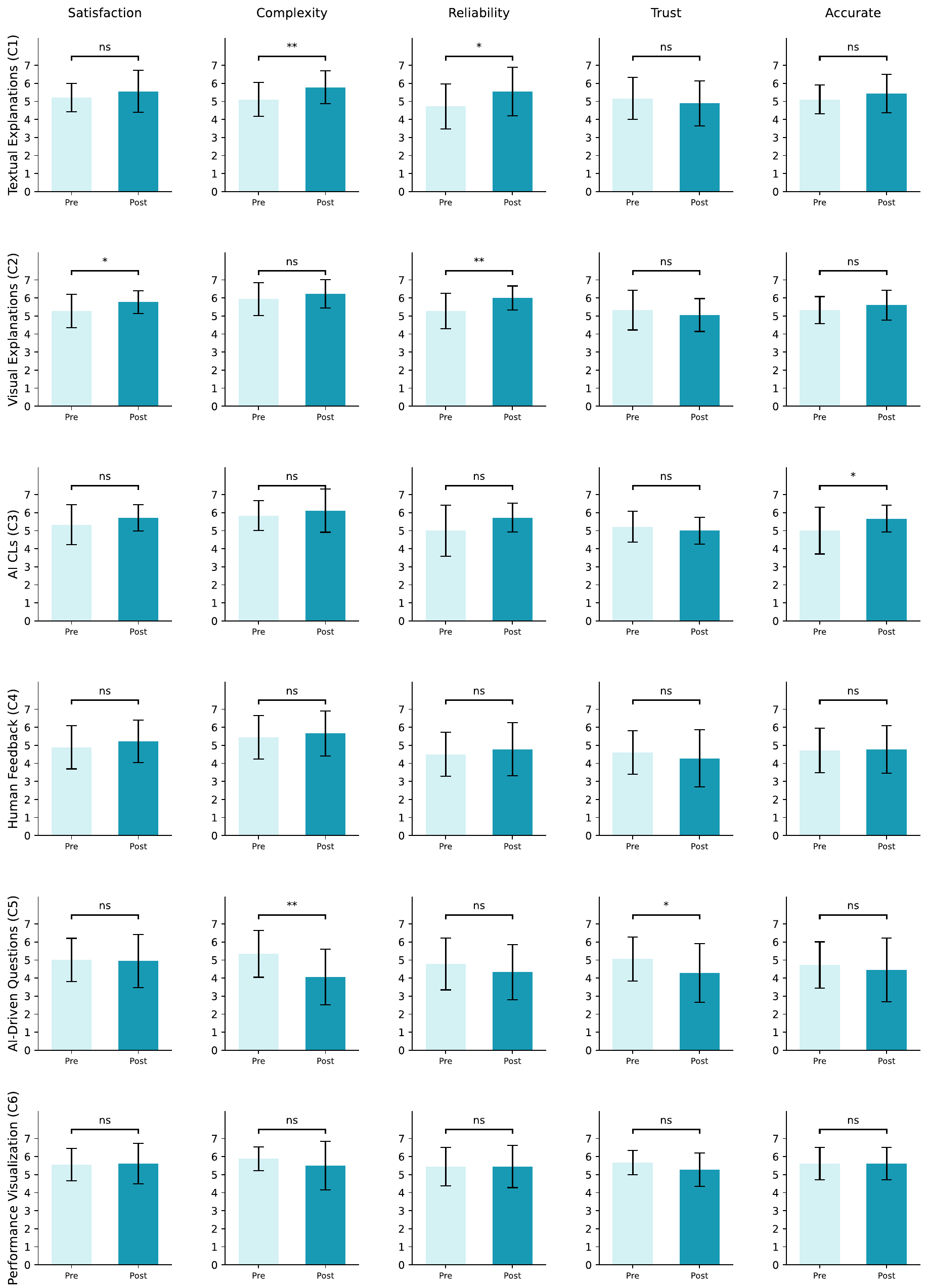}
\caption{Pre- and post-assessment ratings across different criteria: Satisfaction, Complexity, Reliability, Trust, and Accuracy. Each row represents different categories, with error bars indicating variability. Statistical significance is indicated where relevant, with asterisks ($*$) marking significant differences ($p<0.05$) between pre- and post-assessment ratings.}
\label{fig:all_condi}
\end{figure}

\begin{figure}[htbp]
\centering
\includegraphics[width=0.7\textwidth]{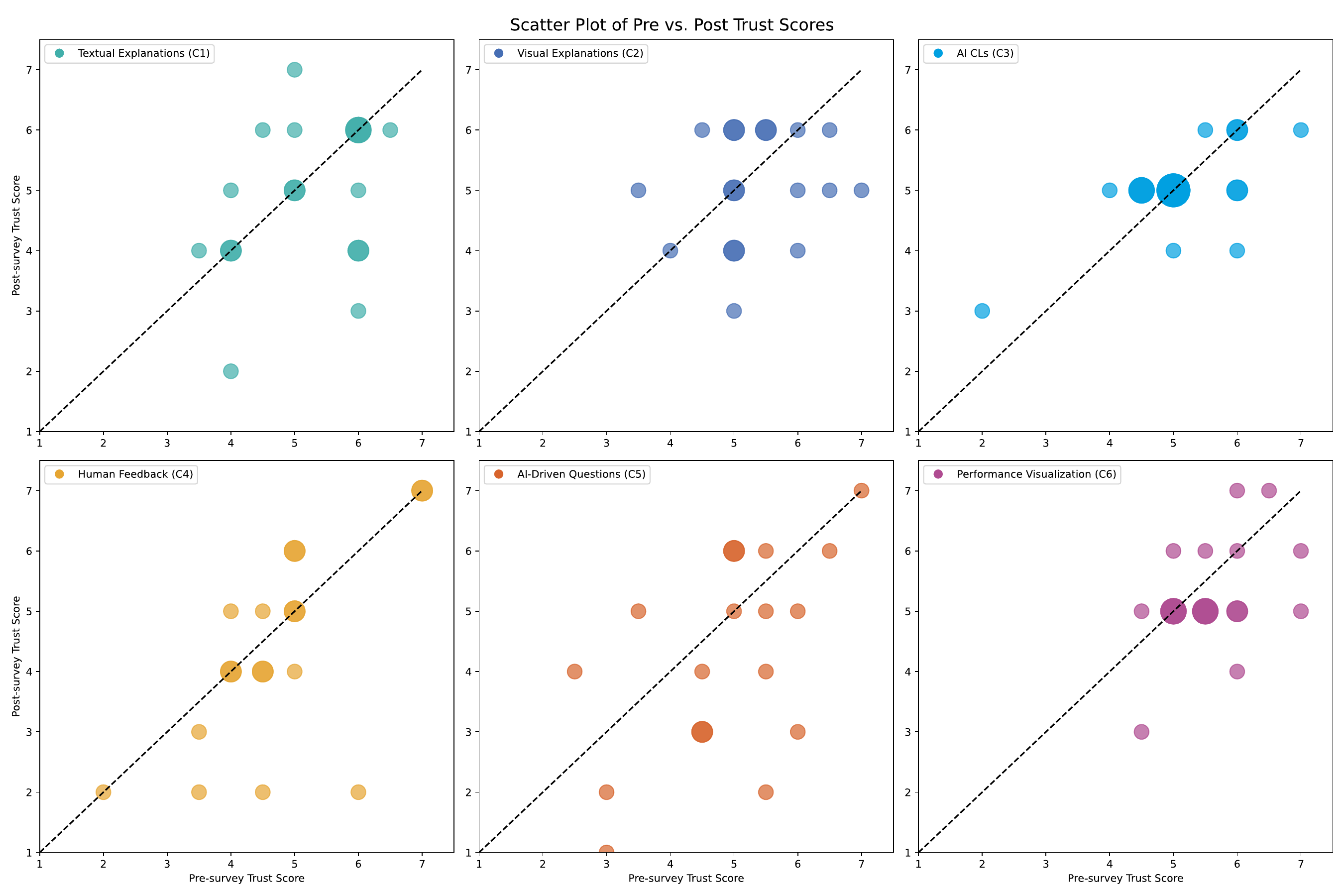}
\caption{Scatter plots illustrating the trust scores of participants across six conditions in pre- and post-questionnaire. The x-axis represents pre-questionnaire trust scores, while the y-axis shows post-questionnaire trust scores. Each point corresponds to a participant, with the size of the point reflecting the number of overlapping participants at a given score. The dashed diagonal line indicates where pre- and post-trust scores are equal, serving as a reference for changes in trust levels after the study.}
\label{fig:pre_post_trust}
\end{figure}

We provide detailed results and statistical analyses to support the findings presented in Section~\ref{sec:results}. Figure~\ref{fig:all_condi} shows the pre- and post-assessment ratings for the five evaluation criteria: Satisfaction, Complexity, Reliability, Trust, and Accuracy. Figure~\ref{fig:mainfig} provides the post-questionnaire scores across six conditions, focusing on the same five measures: satisfaction, system complexity, reliability, trust, and system accuracy. Figure~\ref{fig:pre_post_trust} illustrate changes in participants' self-reported trust between pre- and post-assessments. Table~\ref{tab:within_results} focuses on within-condition comparisons for conditions 1 through 6, reporting the pre- and post-assessment differences for the five evaluation measures. Table~\ref{tab:condition-comparisons} summarizes the statistical comparisons of measures across all six conditions, using the Kruskal-Wallis test statistic to evaluate differences. Table~\ref{tab:engagement-changes} examines engagement changes across session pairs, specifically P1-P2 (P2 Change) and P1-P3 (P3 Change), within each condition. The differences were assessed using the Wilcoxon signed-rank test.

\subsubsection{Pre- and Post-study Questionnaires}

To collect data on participant trust, satisfaction, and engagement with the AI system both before and after the experiment, we conducted a pre-study questionnaire to establish a baseline and a post-study questionnaire to capture the impact of the AI explanation mechanisms. These questionnaires were designed to measure participant familiarity with AI, their prior experience with AI systems, and their perceptions of trust, reliability, and satisfaction during the decision-making process.

\textbf{Pre-Study Questionnaire}

The pre-study questionnaire aimed to collect background information on participant AI usage and their baseline levels of trust in AI systems. It included the following sections:
\begin{enumerate}
    \item \textbf{AI Use:} Participants indicated which AI tools they had previously used (e.g., ChatGPT, Claude.ai, Perplexity) and how frequently they used them, measured on a 7-point Likert scale (1 = not at all, 7 = many times a day).
    \item \textbf{Purpose of AI Use:} Participants were asked to check off all applicable uses for AI, such as writing assistance, information gathering, language translation, or decision-making support.
    \item \textbf{Trust Baseline:} Participants responded to several questions about their satisfaction with AI systems, the complexity of their interactions with AI, and the reliability of AI responses. These items were measured on 7-point scales (e.g., 1 = very dissatisfied, 7 = very satisfied).
    \item \textbf{Knowledge of Diabetes:} To assess familiarity with our specific task, participants rated their knowledge of Type 2 diabetes and its management on a 7-point scale (1 = not familiar at all, 7 = extremely familiar).
\end{enumerate}

\textbf{Post-Study Questionnaire}

After completing the experiment, participants answered the post-study questionnaire. The post-study questionnaire focused on their experience with AI during the meal-planning task. This questionnaire contained both general and condition-specific questions to assess their level of satisfaction, the perceived complexity and reliability of the AI's assistance, and the effectiveness of each AI explanation mechanism.

The post-study questionnaire included the following sections:
\begin{enumerate}
    \item \textbf{General Questions:} Participants rated how satisfied they were with the AI's ability to select the appropriate meal and how accurate and reliable they found the AI's recommendations. The questions were measured using a 7-point Likert scale (1 = not at all, 7 = extremely reliable/accurate).
    \item \textbf{Condition-Related Questions:} Each experimental condition had specific questions about the explanation mechanism used in that condition. Participants rated the usefulness of the explanation mechanism and were prompted to explain how it influenced their decision-making process.
    \item \textbf{Perceptions of Trust and Autonomy:} Participants were asked to reflect on how much they trusted the AI's judgment compared to their own (1 = not at all, 7 = a lot) and whether they felt they had agency in making decisions (1 = not at all, 7 = full autonomy).
    \item \textbf{Cognitive Load:} Participants rated the mental effort required to complete the task and how much the AI reduced the cognitive complexity of the task.
    \item \textbf{Trust Dynamics:} Participants were also asked how their trust in the AI changed from the beginning to the end of the task (1 = no change, 7 = changed completely).
\end{enumerate}

\subsubsection{Questionnaire Content}

\textbf{Pre-Study Questionnaire:}\\

\textbf{[AI Use]}

\begin{itemize}
    \item \textbf{Which AI do you use or have you used (e.g., ChatGPT, Claude.ai, etc.)?} \\
    \textit{Check all that apply:} \\
    ChatGPT, Claude.ai, Perplexity, Pi, Other (please specify)
    
    \item \textbf{How frequently do you use AI?} \\
    \textit{Scale:} 1 (Not at all) to 7 (Many times a day)
\end{itemize}

\textbf{[Purpose of AI Use]}

\begin{itemize}
    \item \textbf{If you use AI, what do you usually use it for?} \\
    \textit{Check all that apply:}
    \begin{itemize}
        \item Writing assistance and editing
        \item Research and information gathering
        \item Problem-solving and brainstorming
        \item Language translation and explanation
        \item Code writing and debugging
        \item Math problem-solving and calculations
        \item Summarizing long texts or articles
        \item Explaining complex concepts in simpler terms
        \item Answering questions on a wide range of topics
        \item Providing general advice on a variety of subjects
        \item Help with making decisions
        \item Other (please specify)
    \end{itemize}
\end{itemize}

\textbf{[Trust Baseline]}

\begin{itemize}
    \item \textbf{How familiar are you with type 2 diabetes and its management?} \\
    \textit{Scale:} 1 (Not familiar at all) to 7 (Extremely familiar)
    
    \item \textbf{How satisfied have you been with the AI's responses in general?} \\
    \textit{Scale:} 1 (Very dissatisfied) to 7 (Very satisfied)
    
    \item \textbf{How easy was it to interact with the AI of your choice?} \\
    \textit{Scale:} 1 (Very complex) to 7 (Very simple)
    
    \item \textbf{How easy was it to understand its response?} \\
    \textit{Scale:} 1 (Very hard) to 7 (Very easy)
    
    \item \textbf{How reliable have you found the AI's responses to be?} \\
    \textit{Scale:} 1 (Not reliable at all) to 7 (Extremely reliable)
    
    \item \textbf{How effectively does the AI understand and respond to your complex queries?} \\
    \textit{Scale:} 1 (Very ineffective) to 7 (Very effective)
    
    \item \textbf{Do you think the AI provides accurate and relevant information in response to your queries?} \\
    \textit{Scale:} 1 (Not accurate at all) to 7 (Extremely accurate)
    
    \item \textbf{How likely are you to use or continue using AI-assisted decision-making?} \\
    \textit{Scale:} 1 (Very unlikely) to 7 (Very likely)
\end{itemize}

\textbf{Post-Study Questionnaire}\\

\textbf{[Satisfaction]}

\begin{itemize}
    \item \textbf{How satisfied were you with our AI's ability to pick the right meal?} \\
    \textit{Scale:} 1 (Very dissatisfied) to 7 (Very satisfied)
\end{itemize}

\textbf{[Complexity]}

\begin{itemize}
    \item \textbf{How clear and understandable was the AI's assistance in helping you choose a meal?} \\
    \textit{Scale:} 1 (Very complex) to 7 (Very simple)
\end{itemize}

\textbf{[Reliability]}

\begin{itemize}
    \item \textbf{How reliable did you find the AI's assistance?} \\
    \textit{Scale:} 1 (Not at all reliable) to 7 (Extremely reliable)
\end{itemize}

\textbf{[Accuracy]}

\begin{itemize}
    \item \textbf{Does the AI provide accurate and relevant information based on your experience or knowledge of meals that type 2 diabetics should eat?} \\
    \textit{Scale:} 1 (Not accurate at all) to 7 (Extremely accurate)
\end{itemize}

\textbf{[Condition-Specific Questions]}

\textbf{Participants answered the following questions based on their assigned condition:}\\

\begin{itemize}

    \item \textbf{Textual Explanation (C1)}:
    \begin{itemize}
        \item \textbf{To what extent did the AI's explanation help you in making the meal decision?} \\
        \textit{Scale:} 1 (Not useful at all) to 7 (Extremely useful)
        \item \textbf{How did the AI's explanation help in your decision-making process?} \\
        \textit{Open text entry}
    \end{itemize}
    
    \item \textbf{Visual Explanation (C2)}:
    \begin{itemize}
        \item \textbf{To what extent did the visual explanation help you in making the meal decision?} \\
        \textit{Scale:} 1 (Not useful at all) to 7 (Extremely useful)
        \item \textbf{How did the visual explanation help in your decision-making process?} \\
        \textit{Open text entry}
    \end{itemize}

    \item \textbf{AI CLs (C3)}:
    \begin{itemize}
        \item \textbf{To what extent did the AI-estimated confidence levels (CLs) help you in making the meal decision?} \\
        \textit{Scale:} 1 (Not useful at all) to 7 (Extremely useful)
        \item \textbf{I thought the estimation of my confidence level was useful.} \\
        \textit{Scale:} 1 (Strongly disagree) to 7 (Strongly agree)
        \item \textbf{How did seeing the confidence levels help in your decision-making process?} \\
        \textit{Open text entry}
    \end{itemize}
    
    \item \textbf{Human Feedback (C4)}:
    \begin{itemize}
        \item \textbf{To what extent did entering confidence levels (CLs) help you in making the meal decision?} \\
        \textit{Scale:} 1 (Not useful at all) to 7 (Extremely useful)
        \item \textbf{How did thinking and inputting the confidence levels help in your decision-making process?} \\
        \textit{Open text entry}
    \end{itemize}
    
    \item \textbf{AI-Driven Questions (C5)}:
    \begin{itemize}
        \item \textbf{To what extent did the AI's questions help you in making the meal decision?} \\
        \textit{Scale:} 1 (Not useful at all) to 7 (Extremely useful)
        \item \textbf{How did you use the AI's questions in your decision-making process?} \\
        \textit{Open text entry}
    \end{itemize}

    \item \textbf{Behavior Visualization (C6)}:
    \begin{itemize}
        \item \textbf{To what extent did the bar chart help you in making the meal decision?} \\
        \textit{Scale:} 1 (Not useful at all) to 7 (Extremely useful)
        \item \textbf{How did you use the bar chart in your decision-making process?} \\
        \textit{Open text entry}
    \end{itemize}

\end{itemize}

\textbf{[General Post-Study Questions for All Conditions]}

We designed the following questions to assess participant experience with the explanation mechanisms across all conditions, including for confidence and clarity, consistency and cognitive load, ease and usefulness, trust and comfort, and trust dynamics.\\

\textbf{Confidence and Understanding:}
\begin{itemize}
    \item \textbf{Confidence in AI's Decision:} \\
    How confident did you feel in the AI's meal choice? \\
    \textit{Scale:} 1 (Not confident at all) to 7 (Extremely confident)
    
    \item \textbf{Understanding of AI Output:} \\
    How well did you understand the [condition-specific content] for picking a meal? \\
    \textit{Scale:} 1 (Not at all) to 7 (Extremely well)
\end{itemize}

\textbf{Consistency and Cognitive Load:}
\begin{itemize}
    \item \textbf{AI Decision Consistency:} \\
    How consistent did you feel the AI's meal choices were across different image pairs? \\
    \textit{Scale:} 1 (Not consistent at all) to 7 (Extremely consistent)
    
    \item \textbf{Mental Demand of Task:} \\
    I thought the task of picking the right meal was mentally demanding. \\
    \textit{Scale:} 1 (Not at all) to 7 (Extremely demanding)
\end{itemize}

\textbf{Ease and Usefulness:}
\begin{itemize}
    \item \textbf{AI's Helpfulness in Decision-Making:} \\
    Did the AI make the task of picking the right meal easier? \\
    \textit{Scale:} 1 (Not at all) to 7 (Very much so)
    
    \item \textbf{Willingness to Use AI System:} \\
    I would like to use this system frequently. \\
    \textit{Scale:} 1 (Strongly disagree) to 7 (Strongly agree)
    
    \item \textbf{Future Use of AI System:} \\
    I would like to use this system in future decision-making tasks. \\
    \textit{Scale:} 1 (Not at all) to 7 (Very much so)
\end{itemize}

\textbf{Trust and Comfort:}
\begin{itemize}
    \item \textbf{Trust in AI vs. Own Judgment:} \\
    How much did you trust the AI's judgment compared to your own? \\
    \textit{Scale:} 1 (Not at all) to 7 (A lot)
    
    \item \textbf{Trust Increase Through AI Assistance:} \\
    Did the AI's assistance increase your trust in its recommendations? \\
    \textit{Scale:} 1 (Not at all) to 7 (Very much so)
    
    \item \textbf{Comfort in Relying on AI:} \\
    How comfortable would you be relying on this AI to help make meal-related decisions for Type 2 diabetics? \\
    \textit{Scale:} 1 (Not at all) to 7 (Extremely comfortable)
\end{itemize}

\textbf{Trust Dynamics:}
\begin{itemize}
    \item \textbf{Trust Change Over Time:} \\
    How did your trust in the AI change from the beginning to the end of the task? \\
    \textit{Scale:} 1 (No change) to 7 (Changed completely)
\end{itemize}

\subsection{Explanation Information Load Quantification for AI Explanation Mechanisms}\label{sec:EPE}

Explanation Information Load (EIL) is designed to quantify the information load imposed on users as they interpret and evaluate the explanation provided by the AI system. The goal is to assess how different types of explanations and decision support mechanisms impact user engagement, trust, and decision-making by causing the user to switch between System 1 and System 2 thinking.

We calculate the EIL based on the amount of information users need to process under each condition basing our approach on Information Theory~\cite{shannon1948mathematical}. The information theory framework can be adapted to count-based metrics, where the amount of information is quantified by the number of distinct elements in the explanation. For example, a logarithmic approach has been successfully employed to quantify the complexity of user interfaces~\cite{10.1145/3025453.3025524,10.1145/3290605.3300866}. 
Higher amounts of information requires users to engage more deeply with the explanations. The processing effort varies depending on the complexity and detail of the designed explanation mechanisms.

\subsubsection{Condition 1: Textual Explanations}
Users are presented with a textual explanation that provides the rationale behind the AI's decision. The EIL for textual explanations is calculated based on the length of the text and the complexity of the language used. We use simple and straightforward language to minimize the complexity. The total EIL for textual explanations is calculated with two components: the information of the text ($I_{text}$) and the information content of the original images ($I_{image}$). We use simple and straightforward language to minimize complexity. The EIL for textual explanations is calculated as follows:
\begin{equation}
    EIL_{textual} = I_{text} + I_{image} 
                = \log_2 (L_{text}) + I_{image}
\end{equation}
where $L_{text}$ is the length of the text.

\subsubsection{Condition 2: Visual Explanations}
Users are presented with segmented images where specific food elements on the meal image are highlighted. Each segment has a corresponding label that explains the content of that segment. 

The total EIL for visual explanations is calculated with three components: the information in the segments ($I_{seg}$) and labels ($I_{label}$), and the information content of the original images ($I_{image}$). The EIL for visual explanations is calculated as follows: 

The EIL for visual explanations is calculated as follows:
\begin{equation}
    EIL_{visual} = \sum (I_{seg} + I_{label} ) + I_{image}
                = \sum ( \log_2 (N_{seg} + L_{label} ) + I_{image}
\end{equation}
where $N_{seg}$ is the number of segments, $L_{label}$ is the length of the label, and $I_{image}$ is the information content of the original images.

\subsubsection{Condition 3: AI CLs}
Users are presented with CLs for both AI and themselves. The EIL for CLs is calculated based on the number of CLs presented ($N_{CL}$). The total EIL for AI CLs is calculated with two components: the information in the CLs ($I_{CL}$) and the information content of the original images ($I_{image}$). The EIL for AI CLs is calculated as follows:
\begin{equation}
    EIL_{CLs} = I_{CL} + I_{image} 
                = \log_2 (N_{CL}) + I_{image}
\end{equation}
where $N_{CL}$ is the number of CLs.

\subsubsection{Condition 4: Human Feedback}
Users input their thinking about the AI's decision and their own CLs. The EIL for human feedback is calculated based on the amount of information users need to input and compare. The total EIL for human feedback is calculated with two components: the information of input ($I_{input}$) and the information content of the original images ($I_{image}$). The EIL for human feedback is calculated as follows:
\begin{equation}
    EIL_{feedback} = I_{input} + I_{image} 
                = \log_2 (N_{input} \times N_{CL}) + I_{image}
\end{equation}
where $N_{input}$ is the number of inputs, and $N_{CL}$ is the number of CLs.

\subsubsection{Condition 5: AI-Driven Questions}
Users are asked three questions to reflect on their decision-making process and the AI's suggestions. The EIL for AI-driven questions is calculated based on the three components: the information in the questions ($I_{ques}$) and the information content of the original images ($I_{image}$). The EIL for AI-driven questions is calculated as follows:
\begin{equation}
    EIL_{questions} = I_{ques} + I_{image} 
                = \sum (\log_2 (L_{ques})) + I_{image}
\end{equation}
where $L_{ques}$ is the length of the questions. We use simple and straightforward language to minimize complexity and ensure consistent language use across all mechanisms. The complexity of the questions is higher than that of Condition 1 (textual explanation), as the questions are generated based on the AI's textual explanation and are designed to enforce users to reflect on AI suggestions. 

\subsubsection{Condition 6: Performance Visualization}
Users are shown a performance visualization chart comparing the AI's performance over its past decisions, in the current task, with their own. The EIL for performance visualization is calculated based on the amount of information users need to process. The total EIL for task performance visualization is calculated with two components: the information of the chart ($I_{chart}$) and the information content of the original images ($I_{image}$). The EIL for behavior visualization is calculated as follows:
\begin{equation}
    EIL_{visualization} = I_{chart} + I_{image} 
                = \log_2 (N_{chart}) + I_{image}
\end{equation}
where $N_{chart}$ is the number of components in the chart. Our chart includes correctness and incorrectness bars for both AI and users. 

\subsubsection{EIL Comparison}
To compare the EIL across the six AI explanation mechanisms, we focus on the relative contributions of different information components while controlling for shared elements. Specifically, since all mechanisms include $I_{image}$, we exclude it from the comparison to isolate the unique contributions of the mechanism-specific components.

\subsection{Calculations} \label{calculation}

\subsubsection{Human CL Prediction for Meal Selection}

In condition 3 (Confidence Levels), participants were provided with CLs for both the AI and themselves. We utilize a generative AI model to predict the participant's CL in selecting the correct meal for controlling blood sugar in diabetes management. The CL represents the confidence level, expressed as a percentage, ranging from 0\% to 100\%. 
In our experiments, we use \texttt{GPT-4o-2024-05-13} as the generative AI model. The model processes historical decision data and the nutritional profiles of the current query meals to predict the user's likelihood of selecting the correct meal. The algorithm for predicting the human CL is shown in Algorithm~\ref{alg:human_cl_prediction}.

\begin{algorithm}[ht]
    \caption{Human Confidence Level Prediction for Meal Selection}
    \label{alg:human_cl_prediction}
    \KwIn{
        Historical decision data $\mathcal{D} = \{D_1, D_2, \dots, D_n\}$, where each $D_i$ contains:
        \begin{itemize}
            \item Nutritional profiles of two meals: $M_{1i}$ (Meal 1), $M_{2i}$ (Meal 2)
            \item Ground truth for correct meal $GT_i$
            \item User's choice $U_i$
        \end{itemize}
        Nutritional profiles for current query meals $M_1$, $M_2$, ground truth $GT_{\text{new}}$
    }
    \KwOut{Predicted confidence level (CL) of the user selecting the correct meal as a percentage (0-100\%)}
    
    \SetKwBlock{Init}{Initialize}{end}
    \SetKwBlock{HistData}{Process Historical Data}{end}
    \SetKwBlock{NewQuery}{Process New Query}{end}
    \SetKwBlock{Predict}{Generate Prediction}{end}
    
    \Init{
        Initialize system prompt $\mathcal{P}_{sys}$: \\
        \hspace{1cm} "Return the predicted likelihood as a percentage (0-100\%) of the user choosing the correct meal based on their past behavior and the nutritional profile required for controlling blood sugar in diabetes patients." \\
        Initialize empty user data prompt $\mathcal{P}_{data}$
    }
    
    \HistData{
        \ForEach{$D_i \in \mathcal{D}$}{
            Extract nutritional profiles $M_{1i}$ and $M_{2i}$ from $D_i$: \\
            \hspace{1cm} $M_{1i} \gets \{calories, fat, carbohydrates, protein\}$ \\
            \hspace{1cm} $M_{2i} \gets \{calories, fat, carbohydrates, protein\}$
    
            Determine ground truth $GT_i$ for $D_i$\;
            
            Record the user's decision $U_i$ for $D_i$\;
    
            Append to data prompt $\mathcal{P}_{data}$: \\
            \hspace{1cm} "Record $i$: $M_{1i}$, $M_{2i}$, Ground Truth = $GT_i$, User's Choice = $U_i$."
        }
    }
    
    \NewQuery{
        Extract nutritional profiles for current query: \\
        \hspace{1cm} $M_1 \gets \{calories, fat, carbohydrates, protein\}$ \\
        \hspace{1cm} $M_2 \gets \{calories, fat, carbohydrates, protein\}$
    
        Determine ground truth $GT_{\text{new}}$ for current query\;
    
        Append current query information to $\mathcal{P}_{data}$: \\
        \hspace{1cm} "New Query: $M_1$, $M_2$, Ground Truth = $GT_{\text{new}}$."
    }
    
    \Predict{
        Send system prompt $\mathcal{P}_{sys}$ and data prompt $\mathcal{P}_{data}$ to a LLM $\mathcal{L}$\;
        
        $\mathcal{L}$ processes historical decisions $\mathcal{D}$ and the nutritional profiles of both current query meals to predict the user's likelihood of selecting the correct meal\;
    
        Return predicted CL as a percentage (0-100\%).
    }
    
    \end{algorithm}

The process involves the following steps:
\begin{enumerate}
    \item Historical Data Compilation: We collect a dataset of the participant's prior decisions $\mathcal{D}$, which consists of $20$ instances where the participant has selected between two meals ($M_1$, $M_2$) with AI assistance but without any explanation (P2). Each instance includes the nutritional profiles for both meals (calories, fat, carbohydrates, protein), the ground truth indicating the correct meal based on blood sugar control guidelines, and the participant's actual choice. This history record serves as the basis for understanding the participant's past behavior and decision patterns.
    \item Prompt Construction: The AI model is initialized with a task-specific prompt that instructs it to return a predicted likelihood of the participant selecting the correct meal. The prompt includes historical data records, where each record contains:
    \begin{enumerate}
        \item Nutritional information for both meals ($M_{1i}$, $M_{2i}$).
        \item Ground truth indicating which meal is optimal for blood sugar control ($GT_i$).
        \item The participant's choice ($U_i$).
    \end{enumerate}

    \item New Query Processing: For the current decision, the nutritional profiles of two new meal options ($M_1$, $M_2$) are appended to the prompt. The correct meal is provided as ground truth ($GT_{\text{new}}$).
    
    \item Prediction: The AI model processes the historical decision data and the nutritional profiles of the new meals. By identifying patterns in how the participant has historically selected meals based on their nutritional content, the model calculates the probability that the participant will make the correct choice in the new task.

\end{enumerate}

\subsubsection{Ground-Truth Calculation}\label{sec:gt}

Effective management of blood sugar levels for diabetes people requires careful selection of meals based on macronutrient composition. In our work, we compute the GT for meal selection by using a set of nutritional guidelines derived from the American Diabetes Association's recommendations for optimal blood sugar control~\cite{colberg2016physical,evert2019nutrition}.

\paragraph{Macronutrient Percentage Calculation}

For each meal, we calculate the percentage contributions of protein, fat, and carbohydrates to the total caloric content. The calculations are based on standard nutritional metrics, using the following formulas:
\begin{equation}
    \text{Protein Percentage (PP)} = \left(\frac{\text{Protein (g)} \times 4}{\text{Total Calories}}\right) \times 100,
\end{equation}

\begin{equation}
    \text{Fat Percentage (FP)} = \left(\frac{\text{Fat (g)} \times 9}{\text{Total Calories}}\right) \times 100,
\end{equation}

\begin{equation}
    \text{Carbohydrate Percentage (CP)} = \left(\frac{\text{Carbohydrates (g)} \times 4}{\text{Total Calories}}\right) \times 100.
\end{equation}

\paragraph{Nutritional Balance Criteria}

Once the macronutrient percentages are determined, we apply a set of balance criteria to assess the suitability of the meal for blood sugar control. Based on dietary guidelines~\cite{evert2019nutrition}, we define balanced ranges for protein and fat percentages:
\begin{equation}
    20\% \leq \text{PP} \leq 32\%,
\end{equation}

\begin{equation}
    20\% \leq \text{FP} \leq 35\%.
\end{equation}

These thresholds are selected to optimize metabolic control, as meals with protein and fat percentages outside these ranges may lead to suboptimal blood glucose regulation.

\paragraph{Balance Indicator}
To simplify the evaluation of each meal, we introduce a binary balance indicator $B$, which serves as a high-level measure of whether the meal meets both the protein and fat balance criteria:

\begin{equation}
    B = 
    \begin{cases} 
      1 & \text{if } 20\% \leq \text{PP} \leq 32\% \text{ and } 20\% \leq \text{FP} \leq 35\%, \\
      0 & \text{otherwise}.
    \end{cases}
\end{equation}

The value of $B=1$ indicates that the meal is balanced with respect to both protein and fat, making it a favorable option for blood sugar management. Conversely, $B=0$ implies that the meal falls outside the desired macronutrient ranges.

\paragraph{Decision Logic}
The final ground-truth decision is made by comparing the balance indicators ($B_1$ and $B_2$) of the two meal options. 
The ground-truth meal selection is made based on the following decision logic~\cite{colberg2016physical,evert2019nutrition}:

\begin{align}
    1. & \quad \text{If } B_1 = 1 \text{ and } B_2 = 0, \text{ choose Meal 1,} \\
    2. & \quad \text{If } B_1 = 0 \text{ and } B_2 = 1, \text{ choose Meal 2,} \\
    3. & \quad \text{If } B_1 = B_2, \text{ choose the meal with the lower CP.}
\end{align}

The decision logic ensures that when both meals satisfy the balance criteria, the meal with the lower carbohydrate content is prioritized, given the direct impact of carbohydrates on postprandial glucose levels.

\bibliographystyle{unsrt}  
\bibliography{reference.bib}

\end{document}